\documentclass[twoside,11pt]{article}

\usepackage{blindtext}

\usepackage[preprint]{jmlr2e}

\usepackage{graphics}
\usepackage{amssymb}
\usepackage{amsmath}
\usepackage{amstext}
\usepackage{subcaption}
\usepackage{threeparttable}
\usepackage{color}
\def\boxit#1{\vbox{\hrule\hbox{\vrule\kern6pt
			\vbox{\kern6pt#1\kern6pt}\kern6pt\vrule}\hrule}}

\def\squarebox#1{\hbox to #1{\hfill\vbox to #1{\vfill}}}
\def\boxit#1{\vbox{\hrule\hbox{\vrule\kern6pt
			\vbox{\kern6pt#1\kern6pt}\kern6pt\vrule}\hrule}}

\makeatletter
\def\namedlabel#1#2{\begingroup
    #2%
    \def\@currentlabel{#2}%
    \phantomsection\label{#1}\endgroup
}
\makeatother

\DeclareMathOperator{\argmin}{argmin}
\DeclareMathOperator{\argmax}{argmax}

\DeclareMathOperator{\Cov}{Cov}

\DeclareMathOperator{\var}{var}

\renewcommand{\tilde}{\widetilde}
\renewcommand{\hat}{\widehat}

\newtheorem{theo}{Theorem} 
\newtheorem{prop}{Proposition} 
\newtheorem{lem}{Lemma}

\graphicspath{{./Figures/}}
\usepackage{lastpage}
\jmlrheading{26}{2025}{1-\pageref{LastPage}}{3/23; Revised
7/24}{5/25}{23-0381}{Ruoxu Tan, Wei Huang, Zheng Zhang and Guosheng Yin}

\ShortHeadings{Functional Treatment Effect}{Tan, Huang, Zhang and Yin}
\firstpageno{1}

\begin{document}

\title{Causal Effect of Functional Treatment}

\author{\name Ruoxu Tan \email ruoxut@tongji.edu.cn \\
       \addr School of Mathematical Sciences and School of Economics and Management\\ 
       Tongji University\\
       Shanghai, China
       \AND
       \name Wei Huang$^{*}$\email wei.huang@unimelb.edu.au \\
       \addr School of Mathematics and Statistics\\
        University of Melbourne\\
       Melbourne, VIC, Australia
       \AND
        \name Zheng Zhang\thanks{Wei Huang and Zheng Zhang are corresponding authors.}  \email zhengzhang@ruc.edu.cn \\
       \addr Center for Applied Statistics, Institute of Statistics $\&$ Big Data\\
       Renmin University of China\\
       Beijing, China
       \AND
       \name Guosheng Yin \email gyin@hku.hk \\
       \addr Department of Statistics and Actuarial Science\\
       University of Hong Kong\\
       Hong Kong SAR, China
  }

\editor{Eric Laber}

\maketitle

\begin{abstract}%   <- trailing '%' for backward compatibility of .sty file
We study the causal effect with a functional treatment variable, where practical applications often arise in neuroscience, biomedical sciences, etc. Previous research concerning the effect of a functional variable on an outcome is typically restricted to exploring correlation rather than causality. %relationship. 
The generalized propensity score, which is often used to calibrate the selection bias, is not directly applicable to a functional treatment variable due to a lack of definition of probability density function for functional data. We propose three estimators for the average dose-response functional based on the functional linear model, 
namely, the functional stabilized weight estimator, the outcome regression estimator and the doubly robust estimator, each of which has its own merits. We study their theoretical properties, which are corroborated through extensive numerical experiments. A real data application on electroencephalography data and disease severity demonstrates the practical value of our methods.
\end{abstract}

 
\section{Introduction}
In observational studies, a fundamental question is to investigate the causal effect of a treatment variable on an outcome. Much of the literature on estimating causal effects from observational data focus on a binary treatment variable (i.e., treatment versus control); see e.g., \citet{Rosenbaum1983,Hirano2003,Imai2014,Chan2016,Athey2018,Ding2019,Tan2020,Guo2021,lin2021estimation}. Recently, there is a growing interest in studying more complex treatment variables, e.g., categorical \citep{Schulte2014,Lopez2017,Lopez2017a,Chen2018,Li2019,Luckett2019} or continuous treatment variables~\citep{Hirano2004,Moodie2014,Galvao2015,Laber2015,Kennedy2017,Fong2018,Kennedy2019,Ai2021,Bonvini2022}.

\begin{figure}[t]
\begin{center}
\includegraphics[width=.5\textwidth]{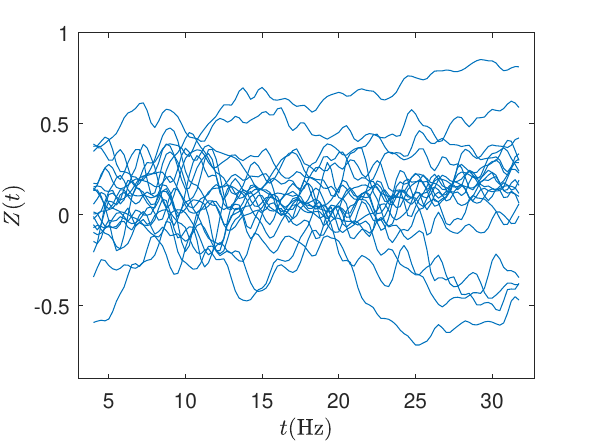}
\caption{A random subsample of 20 frontal asymmetric curves $Z(\cdot)$ on a given frequency domain from the EEG dataset.}\label{fg_sample}
\end{center}
\end{figure}

Our interest is in estimating the average dose-response functional (ADRF) of a functional treatment variable. Functional data are collected repeatedly over a continuous domain, and are fundamentally different from scalar variables due to their infinite dimensionality \citep{Wang2016}. %refer to \citet{Wang2016} for a review on functional data analysis. 
We focus on the causal relationship between a functional treatment variable and a scalar outcome variable in cross-sectional studies, where the treatment variable is real-valued over a continuous domain (e.g., space or frequency, rather than time)
%that does not represent time but some space, frequency, etc., 
and the outcome variable does not vary with respect to that continuous domain. In particular, our work is motivated by an electroencephalography (EEG) dataset~\citep{Ciarleglio2022}, 
 which was collected via a standard headcap with multiple scalp electrodes. The main purpose there is to assess the EEG data as a potential moderator of treatment effect in a clinical trial. Here, we investigate possible causation between the EEG data and the severity of major depressive disorder. A random subsample of 20 frontal asymmetric curves, measuring the difference between the intensities of the human neuronal activities from two electrodes (right and left) placed on the scalp, is shown in Figure~\ref{fg_sample}. For additional applications, see \citet{Morris2015}. As an illustration, we may be interested in the causal relationship between spectrometric data and fat content of certain types of food \citep{Ferraty2006} and the causal effect of body shape represented by the circumference over a given range on human visceral adipose tissue \citep{Zhang2021}. These problems are distinct from sequential decision problems in which multiple treatments are assigned longitudinally \citep{Robins2000a,Robins2000,Moodie2012,Zhao2015,Kennedy2019}. In addition, functional data have been used as one of the covariates in estimating optimal treatment regimes \citep{McKeague2014,Ciarleglio2015,Ciarleglio2016,Ciarleglio2018,Laber2018,Li2023,Park2023}.

To identify the causal effect in an observational study, certain assumptions are required. In addition to standard assumptions such as no interference, consistency and positivity, a routinely used condition for identification is the unconfoundedness assumption \citep{Rosenbaum1983}, i.e., the set of potential outcomes is conditionally independent of the treatment given the observed confounders. For a continuous scalar treatment, under the unconfoundedness assumption, the generalized propensity score or the stabilized weight \citep{Hirano2004,Imai2004,Galvao2015,Kennedy2017,Ai2021,Bonvini2022} has been widely used to calibrate the selection bias via estimating equations. It relies upon the (conditional) density of the treatment variable. However, the probability density function of functional data generally does not exist (\citealp*{Delaigle2010}). Therefore, in our context, neither the generalized propensity score nor the stabilized weight are well defined, and thus traditional methods are not directly applicable. To circumvent this issue, \citet{Zhang2021} proposed to approximate the functional treatment variable based on functional principal component (PC) analysis. They then defined a stabilized weight function by the (conditional) probability density of the PC scores to adjust for the selection bias. However, as the number of PCs approaches infinity, this stabilized weight becomes ill-defined, and the consistency of their estimator is not guaranteed.

Under the unconfoundedness assumption in the context of a functional treatment variable, we propose three new estimation approaches to the ADRF: a functional stabilized weight (FSW) method, a partially linear outcome regression (OR) method, and a doubly robust (DR) approach combining the first two. The first method assumes a functional linear model for the ADRF and incorporates a new FSW. Our FSW is well-defined and guarantees the identification of the ADRF for any functional treatment variable. We propose a novel, consistent, nonparametric FSW estimator and estimate the ADRF using functional linear regression with a weighted outcome. The OR method assumes a more restrictive additive model that considers confounding variables. It quantifies the ADRF by the expectation of the parametric model with respect only to the confounding variables, which is computed through a backfitting algorithm. Our DR estimator is consistent if either of the first two estimators is consistent. If the partially linear model is correctly specified, the DR estimator converges at the same fast rate as the OR estimator. Our double robustness is different from those in the literature on a continuous scalar treatment variable~\citep[see e.g.,][]{Kennedy2017,Bonvini2022}, where the estimator is consistent if one of two models for the conditional outcome and the generalized propensity score is correctly specified. In our case, the functional linear model for the ADRF is required for the consistency of all three estimators, while the FSW, serving the role of the generalized propensity score, is estimated nonparametrically.

More recently, \citet{Wang2023} proposed a so-called weight-modified kernel ridge regression (WMKRR) estimator for the ADRF. They assumed that the ADRF lies in a functional reproducing kernel Hilbert space (RKHS) and utilized the kernel ridge regression to estimate the ADRF. Their method also requires an estimation of the FSW. To this end, they proposed to minimize a novel uniform balancing error derived from their estimation of the ADRF. Although the kernel ridge regression is more flexible than the functional linear model that is used in our methods, it still faces the risk of model misspecification since the model depends on the choice of the kernel. We will compare our methods with those from \citet{Zhang2021} and \citet{Wang2023} in simulation experiments.

The rest of the paper is organized as follows. We introduce the basic setup for the identification of ADRF in Section~\ref{sc2}, followed by the development of our three estimators in Section~\ref{sc3}. We investigate the convergence rates of the estimators in Section~\ref{sc4}. The selection of tuning parameters is discussed in Section~\ref{sccom}. Simulation experiments and the real data analysis are presented in Sections~\ref{scSim} and \ref{scReal}, respectively. We conclude in Section~\ref{sc7}. Proofs and additional technical details are provided in the Appendix.

\section{Identification of the Average Dose-Response Functional}\label{sc2}

We consider a functional treatment variable such that the observed treatment variable, denoted by $Z=Z(\cdot):\mathcal{T}\to \mathbb{R}$, is a smooth real-valued random function defined on a compact interval $\mathcal{T}$, with $E\{\int_\mathcal{T} Z^2(t)\,dt\}< \infty$. Let $Y^*(z)$ be the potential outcome given the treatment $Z=z$ for $z\in L^2(\mathcal{T})$, the Hilbert space of squared integrable functions on $\mathcal{T}$. Let ${X}=(X_1,\ldots,X_p)^\top \in \mathbb{R}^p$ be the $p$-dimensional observable confounding covariates that are related to both $Y^*(z)$ and $Z$. 

For each individual, we only observe a confounding vector $X$, a random treatment $Z$ and the corresponding outcome $Y = Y^*(Z) \in \mathbb{R}$. We further assume that the functional variable $Z$ is fully observed. Essentially, our methodology can be applied when $Z$ is only sparsely observed as long as the trajectory of $Z$ can be well reconstructed, while the development of asymptotic properties in such cases would be much more challenging~\citep{Zhang2016,Zhou2022}.

We are interested in estimating the average dose-response functional (ADRF), $E\{Y^*(z)\}$, for any $z\in L^2(\mathcal{T})$. Because we never observe $Y^*(z)$ simultaneously for all $z\in L^2(\mathcal{T})$, 
%it is impossible 
to identify $E\{Y^*(z)\}$ from the observed data,
we impose the following identification assumptions.

\paragraph{Assumption 1}

\begin{itemize}
\item [(i)]	Unconfoundedness:
Given $ {X}$, $Z$ is independent of $Y^*(z)$ for all $z\in L^2(\mathcal{T})$, i.e., $\{Y^*(z)\}_{z\in L^2(\mathcal{T})}\perp Z\mid  {X}$.
\item [(ii)] No interference: There is no interference among the units, i.e., each individual’s outcome depends only on their own treatment.
\item [(iii)] Consistency: $Y=Y^*(z)$ a.s.~if $Z=z$.
\item [(iv)] Positivity: The conditional density 
$f_{ {X}|Z}$ 
satisfies $f_{ {X}|Z}( {X}|Z)>0$ a.s.
\end{itemize}

Assumption 1 is a natural extension of the classical identification assumption in the literature on the scalar treatment effect  
to the context of a functional treatment variable. In particular, we focus on cross-sectional studies where the interval $\mathcal{T}$ does not represent a time interval and there is no dynamic confounding effect. To circumvent the problem of the nonexistent probability density function of functional data, the positivity condition (iv) is imposed on $f_{X|Z}$ instead of $f_{Z|X}$.

Under Assumption 1, the ADRF can be identified from the observable variables $(X,Y,Z)$ in the three ways as follows. First, it is easy to verify
\begin{align}
E\{Y^*(z)\}&=E[E\{Y^*(z)| {X}\}]=E[E\{Y^*(z)| {X},Z=z\}]
=E\{E(Y| {X},Z=z)\}.\label{id:OR}
\end{align}

Second, we can further write
\begin{align}
E\{Y^*(z)\} &=E\{E(Y| {X},Z=z)\}\notag\\
&=\int_{\mathbb{R}^p} E(Y| {X}={x},Z=z) f_{ {X}}( {x})\,d {x}\notag \\
&=\int_{\mathbb{R}^p} E(Y| {X}= {x},Z=z)\dfrac{f_{ {X}}( {x})}{f_{ {X}|Z}( {x}|z)} f_{ {X}|Z}( {x}|z) \,d {x}\notag\\
&= E\{\pi_0(Z, {X}) Y |Z=z \}\label{id:IPW}\,,
\end{align}
where $\pi_0(Z, {X})=f_{X}( {X}) / f_{ {X}|Z}( {X}|Z)$ is called the functional stabilized weight (FSW). If $Z$ is a scalar random variable, we have $f_{X}( {X})/f_{ {X}|Z}( {X}|Z)=f_Z(Z)/f_{Z| {X}}(Z| {X})$, where the right-hand side is the classical stabilized weight.

Third, let $E_{ {X}}$ denote the expectation with respect to ${X}$. Noting that
\begin{align*}
E[E_{ {X}}\{E(Y| {X},Z)\} | Z=z]&=E\{E(Y| {X},Z=z)\}\,, \\
E\{E(Y| {X},Z)\pi_0( {X},Z)|Z=z\}&=E\{Y\pi_0( {X},Z)|Z=z\}\,,
\end{align*}
 together with \eqref{id:OR} and \eqref{id:IPW}, we can also identify $E\{Y^*(z)\}$ as
\begin{align}
E\{Y^*(z)\} = E[\{Y-E(Y| {X},Z)\}\pi_0( {X},Z)+E_{ {X}}\{E(Y| {X},Z)\} | Z=z]\,,\label{id:DR}
\end{align}
because the second and third terms above cancel out.

Based on the three identification strategies in \eqref{id:OR}, \eqref{id:IPW}, and \eqref{id:DR} respectively, we develop three estimators of $E\{Y^*(z)\}$. Each estimator is based on different model assumptions and has its own merits.

\section{Models and Estimation Methods}\label{sc3}
 
For all three methods, we adopt a functional linear model on the ADRF,
\begin{align}\label{eq1}
  E\{Y^*(z)\}= a + \int_\mathcal{T} b(t) z(t)\,dt \,,
  \end{align}
  where $a$ is a scalar intercept, and $b\in L^2(\mathcal{T})$ is a slope function.  The functional linear model 
  %for the observed outcome $Y$ 
  is widely used in 
  %the literature of 
  scalar-on-function regression, whereas the model here is assumed for the potential outcome $Y^*(z)$.
   Essentially,  model~\eqref{eq1} can be replaced by any other nonlinear regression models \citep[e.g.,][]{Muller2008,Li2010,Wang2023} without affecting our main idea. We choose to use the functional linear model mainly due to its simplicity and interpretability. The slope function $b=b(\cdot)$ is a quantity of primary interest convenient for the causal inference because it measures the intensity of the functional treatment effect. For example, to compute the average treatment effect (ATE) between $z_1,z_2\in L^2(\mathcal{T})$, $E\{Y^*(z_1)-Y^*(z_2)\}$, it suffices to compute $\int_\mathcal{T} b(t)\{z_1(t)-z_2(t)\} \,dt$. Additionally, the slope function provides useful causal information (e.g., a positive/negative causal relationship), even if the model is misspecified.

\subsection{Functional Stabilized Weight Estimator}\label{sc32} 
We construct our functional stabilized weight (FSW) estimator of the ADRF based on equation~\eqref{id:IPW} and model~\eqref{eq1}. In this construction, we propose a nonparametric estimator for the FSW. In particular, equation~\eqref{id:IPW} suggests that we can estimate $E\{Y^*(z)\}$ in \eqref{eq1} using the classical functional linear regression technique~\citep[e.g.,][]{Hall2007}. Here, the response variable $Y$ is replaced by $\hat{\pi}(Z, {X})Y$, provided an estimator of $\pi_0$, $\hat{\pi}$.  To facilitate the development, suppose that an estimator $\hat{\pi}$ of $\pi_0$ is available for the time being. Let $G(t,s)=\Cov \{Z(t),Z(s)\}$ be the covariance function of $Z$. Its eigenvalues $\lambda_1\geq \lambda_2\geq \cdots \geq 0$ and the corresponding orthonormal eigenfunctions  $\{\phi_j\}_{j=1}^\infty$ are defined as follows,
\begin{equation}
\int_\mathcal{T} G(t,s)\phi_j(s)\,ds = \lambda_j \phi_j(t) \,, \textrm{ for all $t\in\mathcal{T}$\,,} \label{eq:EigenDecomposition}
\end{equation}
where $\{\phi_j\}_{j=1}^\infty$, also called the PC basis, is a complete basis of $L^2(\mathcal{T})$.
According to the Karhunen--Lo\'eve expansion, we have
\begin{equation}
Z(t) = \mu_Z(t) + \sum_{j=1}^\infty \xi_j \phi_j(t)\,,\label{eq:KLexpansion}
\end{equation}
where $\mu_Z(t)=E\{Z(t)\}$ and $\xi_j=\int_\mathcal{T} \{Z(t)-\mu_Z(t) \} \phi_j(t)\,dt$.

With a slight abuse of notation, let $G:L^2(\mathcal{T})\to L^2(\mathcal{T})$ denote a linear operator such that $(G\circ \psi) (t) = \int_{\mathcal{T}} G(t,s) \psi(s)\,ds$, for a function $\psi \in L^2(\mathcal{T})$. If we let $\mu_{Y,\pi}=E\{\pi_0(Z, {X})Y\}=a+\int_\mathcal{T} b(t) \mu_Z(t)\,dt$, by combining~\eqref{id:IPW}, \eqref{eq1} and \eqref{eq:KLexpansion}, we can write
\begin{align*}
E\{\pi_0(Z, {X}) Y |Z=z \} - \mu_{Y,\pi} = \int_\mathcal{T} b(t)\{z(t)-\mu_Z (t)\}\,dt\,.
\end{align*}
It follows that $(G\circ b)(t)=e(t)$, where $e(t) = E[\{ \pi_0(Z, {X}) Y-\mu_{Y,\pi} \} \{Z(t)-\mu_Z(t) \}]$. Noting that $\{\phi_j\}_{j=1}^\infty$ is an orthonormal complete basis of $L^2(\mathcal{T})$, we can write $b(t)=\sum_{j=1}^\infty b_j\phi_j(t)$ and $e(t)=\sum_{j=1}^\infty e_j\phi_j(t)$, and combining them with \eqref{eq:EigenDecomposition}, we obtain that, for all $j$, 
\begin{align*}
b_j = \int_{\mathcal{T}} b(t)\phi_j(t)\,dt\,, \textrm{ and }
e_j = \int_{\mathcal{T}} e(t)\phi_j(t) \,dt=\int_{\mathcal{T}} (G\circ b)(t)\phi_j(t)\,dt =b_j\lambda_j\,.
\end{align*}

The above arguments suggest we can estimate $b(t)$ by $\hat{b}_{\mathrm{FSW}}(t) = \sum_{j=1}^q \hat{b}_{\mathrm{FSW},j} \hat{\phi}_j(t)$, where
 $q$ is a truncation parameter and $\hat{b}_{\mathrm{FSW},j}=\hat{\lambda}_j^{-1}\hat{e}_{\mathrm{FSW},j}$ is an estimator of $b_j$. The $\hat{\lambda}_j$'s and $\hat{\phi}_j$'s are the eigenvalues and eigenfunctions of the empirical covariance function $\hat{G}(s,t)= \sum_{i=1}^n\{Z_i(s)-\hat{\mu}_Z(s)\}\{Z_i(t)-\hat{\mu}_Z(t)\}/n$, respectively, where $\hat{\mu}_Z(t)=\sum_{i=1}^n Z_i(t)/n$ is the empirical estimator of $\mu_Z(t)$. Moreover, by the definition of $e(t)$, we can estimate it by
$$\hat{e}_{\mathrm{FSW}}(t) = \sum_{i=1}^n \dfrac{\{Y_i\hat{\pi}(Z_i, {X}_i)-\hat{\mu}_{Y,\pi} \} \{Z_i(t)-\hat{\mu}_Z(t)\}}{n}$$
 with $\hat{\mu}_{Y,\pi} = \sum_{i=1}^n  Y_i\hat{\pi}(Z_i, {X}_i)/n$, and take $\hat{e}_{\mathrm{FSW},j} = \int_\mathcal{T} \hat{e}_{\mathrm{FSW}}(t)\hat{\phi}_j(t)\,dt$ as an estimator of $e_j$. Finally, the estimator of the intercept $a$ is given by
 $\hat{a}_\mathrm{FSW}= \hat{\mu}_{Y,\pi} -\int_{\mathcal{T}} \hat{b}_{\mathrm{FSW}}(t)\hat{\mu}_Z(t)\,dt$.

It remains to estimate the FSW, $\pi_0(Z, {X})=f_{X}( {X})/f_{ {X}|Z}( {X}|Z)$. 
%In particular, we need to estimate the weights at the sample points, $\pi_0(Z_i, {X}_i)$, for $i=1,\ldots,n$. 
A naive approach would first estimate the densities $f_{X}$ and $f_{ {X}|Z}$ and then take the ratio. However, this may lead to unstable estimates because an estimator of the denominator $f_{ {X}|Z}$ is difficult to derive and may be too close to zero. Also, approximating $Z$ by its PC scores would not alleviate the challenge but pose additional difficulty in theoretical justification. To circumvent these problems,
we treat $\pi_0$ as a whole and develop a robust nonparametric estimator.  We can find the conditional moments that identify $\pi_0(z,\cdot)$ for every fixed $z$, 
%The idea of estimating the density ratio directly has also been exploited in \cite{Ai2021} for a scalar treatment variable.  However, their method is not applicable to the functional treatment.
%Specifically, when the treatment $Z$ was a scalar, \cite{Ai2021} found that the moment equation
%\begin{equation}
%E\{\pi_0(Z, {X})v( {X})u(Z)\} = E\{u(Z)\}E\{v( {X})\}\,,\label{UCM}
%\end{equation}
%holds for any integrable functions $v( {X})$ and $u(Z)$ and it identifies $\pi_0$. {\color{red} However, in the case of functional treatment, it is unclear whether \eqref{UCM} still identifies $\pi_0(\cdot,\cdot):L^2(\mathcal{T})\times\mathcal{X}\to \mathbb{R}$, where $\mathcal{X}$ denotes the support of $X$. The space of functionals $\{u(\cdot):L^2(\mathcal{T})\to\mathbb{R}\}$ is generally nonseparable, which impedes a consistent approximation of $\pi_0(\cdot,\cdot)$ using the basis expansion.
%} 
%To overcome the challenge, {\color{blue}instead of estimating the function $\pi_0(\cdot,\cdot):L^2(\mathcal{T})\times \mathcal{X}\to\mathbb{R}$, we propose to estimate its projection $\pi_0(z,\cdot):\mathcal{X}\to\mathbb{R}$ for every fixed $z\in L^2(\mathcal{T})$.} 
\begin{align}\label{eq2}
	E\{\pi_0(z, {X})v( {X})|Z=z\}=\int_{\mathbb{R}^p}\dfrac{f_{X}(x)}{f_{X|Z}(x|z)} v(x) f_{X|Z}(x|z) d {x} =E\{v(X)\},
\end{align}
which holds for any integrable function $v(X)$. 
%The following proposition shows the identification.
\begin{prop}\label{prop1}
For any fixed $z\in L^2(\mathcal{T})$,
\begin{align*}
E\{\pi(z,X)v(X)|Z=z\}=E\{v(X)\}\,,
\end{align*}
 holds for all integrable functions $v(X)$ if and only if $\pi(z,X)=\pi_0(z,X)$ a.s.
\end{prop}
The proof is given in Appendix~\ref{Appsc2}. This result indicates that $\pi_0$ is fully characterized by~\eqref{eq2}, and thus we can use it to define our estimator. Since~\eqref{eq2} is defined for a fixed $z$, we need to estimate $\pi_0$ at all sample points $Z_i$,  
where $(X_i,Y_i,Z_i)$, for $i=1,\ldots,n$, are i.i.d.~copies of $(X,Y,Z)$.

We consider the leave-one-out index set $\mathcal{S}_{-i}=\{j:j\neq i,j=1,\ldots,n\}$ to estimate $\pi_0$ at each $Z_i$. Specifically, for a given sample point $z=Z_i$, the right-hand side of~\eqref{eq2} can be estimated by its empirical version $\sum_{j \in \mathcal{S}_{-j}} v(X_j)/(n-1)$.  We propose to estimate the left-hand side of~\eqref{eq2} by the Nadaraya--Watson estimator for a functional covariate~\citep{Ferraty2006},
\begin{align*}
\dfrac{\sum_{j\in \mathcal{S}_{-i}} \pi_0(Z_i,X_j)v(X_j)K\{d(Z_j,Z_i) / h\} }{\sum_{j\in \mathcal{S}_{-i}}  K\{d(Z_j,Z_i)/h\}}=\dfrac{1}{n-1}\sum_{j \in \mathcal{S}_{-j}} v(X_j),
\end{align*}
where $d(\cdot,\cdot)$ denotes a semi-metric in $L^2(\mathcal{T})$, $h>0$ is a bandwidth and $K$ is a kernel function quantifying the proximity of $Z_j$ and $Z_i$. Commonly used choices for $d$ include the $L^2$ norm and the projection  metric $d(z_1,z_2) = [\int_\mathcal{T} \{\Pi(z_1)-\Pi(z_2)\}^2\,dt]^{1/2}$, where $\Pi$ denotes a projection operation to a subspace of $L^2(\mathcal{T})$. 
%The projection metric may yield better asymptotic properties; see Section~\ref{sc4} for more details, and see Section~\ref{sccom} on how to choose $h$ and $K$.

As it is not possible to solve~\eqref{eq2} for an infinite number of $v$'s from a given finite sample, we use a growing number of basis functions to approximate any suitable function $v$. Specifically, let $\nu_{k}(X) = \big(v_{1}(X),\ldots, v_{k}(X)\big)^\top$ be a set of known basis functions, which may serve as a finite-dimensional sieve approximation to the original function space of all integrable functions. We define our estimator of $\pi_0(Z_i,X_j)$, for $j\in \mathcal{S}_{-i}$, as the solution to the  $k$ equations,
\begin{align}\label{eq22}
\dfrac{\sum_{j\in \mathcal{S}_{-i}}  \pi(Z_i,X_j)\nu_k(X_j)K\{d(Z_j,Z_i) / h\} }{\sum_{j \in \mathcal{S}_{-i}}  K\{d(Z_j,Z_i) / h\}} = \dfrac{1}{n-1} \sum_{j\in \mathcal{S}_{-i}} \nu_k(X_j)\,,
\end{align}
which asymptotically identifies $\pi_0$ as $k\rightarrow \infty$ and $h\rightarrow 0$.
However, in practice, a finite number $k$ of equations  in~\eqref{eq22} cannot identify a unique solution. Indeed, for any strictly increasing and concave function $\rho$, replacing $\pi(Z_i,X_j)$ by $\rho'\{ \hat{\eta}_{Z_i}^\top \nu_k(X_j) \}$
would satisfy \eqref{eq22}. Here, $\rho'(\cdot)$ is the first derivative of $\rho(\cdot)$, and $\hat{\eta}_{Z_i}$ is a $k$-dimensional vector that maximizes the strictly concave function $H_{hk,Z_i}$, defined as
\begin{align}\label{eq22.5}
H_{hk,Z_i}(\eta) = \dfrac{\sum_{j\in \mathcal{S}_{-i}} \rho\{ \eta^\top \nu_k(X_j)\} K\{d(Z_j,Z_i) / h\} }{\sum_{j\in \mathcal{S}_{-i}} K\{d(Z_j,Z_i) / h\}} -\eta^\top\bigg\{ \dfrac{1}{n-1} \sum_{j\in \mathcal{S}_{-i}} \nu_k(X_j)\bigg\}\,.
\end{align}
This can be verified by noting that the gradient
$\nabla H_{hk,Z_i} (\hat{\eta}_{Z_i})=0$ corresponds to \eqref{eq22} with $\pi(Z_i,X_j)$ replaced by $ \rho'\{ \hat{\eta}_{Z_i}^\top \nu_k(X_j) \}$. %In other words, for each strictly increasing and concave function $\rho$, $\rho'\{ \hat{\eta}_{Z_i}^\top \nu_k(X_j) \}$ {\color{red}can be} an estimator of $\pi_0(Z_i,X_j)$.
Therefore, for a given strictly increasing and concave function $\rho$ together with $\hat{\eta}_{Z_i}$ defined above, we define our estimator of $\pi_0(Z_i,X_j)$ as $\hat{\pi}_{hk}(Z_i,X_j) =  \rho'\{ \hat{\eta}_{Z_i}^\top \nu_k(X_j) \}$. This also induces the estimator $\hat{\pi}_{hk}(Z_i,X_i) =  \rho'\{ \hat{\eta}_{Z_i}^\top \nu_k(X_i) \}$ with $X_j$ replaced by $X_i$.

Our estimator $\hat{\pi}_{hk}$ has a generalized empirical likelihood interpretation for a variety of choices for $\rho$. Specifically, we show in Appendix~\ref{Appsc1} that the estimator defined by~\eqref{eq22.5} is the dual solution to minimizing some distance measure between the FSW and 1 locally for each $Z_i$, subject to the constraint \eqref{eq22} (i.e., the constraint \eqref{eq24} in Appendix~\ref{Appsc1}).
That is, our estimator $\widehat{\pi}_{hk}$ is the desired weight closest to the baseline uniform weight 1 locally in a small neighbourhood of each $Z_i$, subject to the finite sample version of the moment restriction in Proposition~\ref{prop1}. The uniform weight can be considered as a baseline. This is because, when $Z$ and $Y^*(z)$ are unconditionally independent for all $z\in L^2(\mathcal{T})$ (i.e.,~a randomized trial without confoundedness), the functional stabilized weights should be uniformly equal to 1.
Different choices of $\rho$ correspond to different distance measures (see Appendix~\ref{Appsc1} for other examples).  
We suggest to choose $\rho(v) = -\exp(-v-1)$ which guarantees a positive weight estimator $\widehat{\pi}_{hk}(Z_i,X_i)$. 

\subsection{Outcome Regression Estimator and Doubly Robust Estimator}\label{sc33}
To estimate the ADRF based on the outcome regression (OR) identification method in \eqref{id:OR}, an estimator of $E(Y|{X},Z)$ is required. Thus, based on the functional linear model in \eqref{eq1}, we further assume a partially linear additive model,
\begin{align}\label{MixedLinear}
Y = a + \int_\mathcal{T} b(t)Z(t) \,dt + g({X};\theta) + \epsilon\,,
\end{align}
where $a$ and $b(\cdot)$ are the same as those in~\eqref{eq1}, $g(\cdot;\theta)$ is a known function with an unknown parameter $\theta\in \mathbb{R}^p$ and $\epsilon$ is an error variable with $E(\epsilon|Z,X)=0$ and $E(\epsilon^2|Z,X)<\infty$ uniformly for all $Z$ and $X$. Note that $X$ and $Z$ are generally dependent. 
%In what follows, 
We mainly investigate the linear function $g(X;\theta) = \theta^\top {X}$, where ${X}$ may involve some transformations, e.g., the log-transformation, polynomials, and exponential of some elements of ${X}$. 
%Under such a setting, we call~\eqref{MixedLinear} a partially linear model.
 Without loss of generality, we assume that $E(\theta^\top X)=0$; otherwise the non-zero mean can be absorbed into $a$. Model~\eqref{MixedLinear} may be extended to include interactions between $Z$ and $X$. However, research on functional models involving interactions between functional and multivariate continuous variables is scarce, even in the literature on non-causal functional regression. Such extension is beyond the scope of this paper and deserves future research.

Recalling $E\{Y^*(z)\}=E\{E(Y|X,Z=z)\}$ from \eqref{id:OR}, it is clear that \eqref{MixedLinear} implies \eqref{eq1}, and thus  model~\eqref{MixedLinear} imposes a stronger structure than model~\eqref{eq1}. Specifically, the linear part $\theta^\top X$ can be replaced by any nonparametric function of $X$ whose mean is zero, which still implies model~\eqref{eq1}.
However, model~\eqref{MixedLinear} gives a simpler estimator of $E\{Y^*(z)\}$ with a faster convergence rate and a better interpretability of the effects of the treatment and  confounding variables on the outcome.
%Based on \eqref{id:OR} and model \eqref{MixedLinear}, we deduce that, for all $z\in L^2(\mathcal{T})$,
%\begin{align*}
%E\{Y^*(z)\} =E\{E(Y|X,Z=z)\} = a + \int_\mathcal{T} b(t)z(t) \,dt\,,
%\end{align*}
%the same as~\eqref{eq1}. 

%\begin{rem}
%By the FPCA of $Z$, one can also use a joint least square method to estimate $a$, $b$ and $\theta$ simultaneously as in \citet{Shin2009}, which is theoretically equivalent to the backfitting algorithm. However, we found in numerical studies that this joint estimation procedure produces results inferior to our backfitting algorithm possibly because of the correlation between $Z$ and $X$.
%\end{rem}

The OR estimators $\hat{a}_{\mathrm{OR}}$, $\hat{b}_{\mathrm{OR}}$ and $\hat{\theta}_{\mathrm{OR}}$ can be obtained by adapting the backfitting algorithm~\citep{Buja1989}. In the first step, we set $\theta$ to be zero and apply the same method as in Section~\ref{sc32} to regress $Y$ on $Z$, which gives the initial estimators of $a$ and $b$ in \eqref{MixedLinear}. Specifically, we estimate $b(t)$ initially by $
\hat{b}_{\mathrm{ini}}(t) = \sum_{j=1}^q \hat{b}_{\mathrm{ini},j} \hat{\phi}_j(t)$ with $\hat{b}_{\mathrm{ini},j}=\hat{\lambda}_j^{-1}\hat{e}_{\mathrm{ini},j}$, where $\hat{\lambda}_j$'s and $\hat{\phi}_j$'s are the same as those in Section~\ref{sc32}, while $\hat{e}_{\mathrm{ini}}(t)=\sum_{i=1}^n  (Y_i-\hat{\mu}_Y ) \{Z_i(t)-\hat{\mu}_Z(t)\}/n$ with $\hat{\mu}_Y  = \sum_{i=1}^n  Y_i/n$. We then define $\hat{e}_{\mathrm{ini},j} = \int_\mathcal{T} \hat{e}_{\mathrm{ini}}(t)\hat{\phi}_j(t)\,dt$. Finally, the initial estimator $\hat{a}_\mathrm{ini}$ of $a$ is given by $ \hat{\mu}_Y -\int_{\mathcal{T}} \hat{b}_{\mathrm{ini}}(t)\hat{\mu}_Z(t)\,dt$.

In the second step, we use the traditional least square method to regress the residual $Y_i-\hat{a}_{\mathrm{ini}}-\int_{\mathcal{T}}\hat{b}_{\mathrm{ini}}(t)Z_i(t)\,dt$ on $X_i$, for $i=1,\ldots,n$. Specifically, we estimate $\theta$ by $\hat{\theta}_{\mathrm{ini}} = (\Xi^\top \Xi)^{-1} \Xi^\top \mathbb{Y}_{\mathrm{res}}$, where $\Xi$ is an $n\times p$ matrix with the $(i,j)$-th entry being $X_{ij}$, the $j$-th element of $X_i$, and
$$ \mathbb{Y}_{\mathrm{res}}=\Big\{Y_1-\hat{a}_{\mathrm{ini}}-\int_{\mathcal{T}}\hat{b}_{\mathrm{ini}}(t)Z_1(t)\,dt,\ldots,Y_n-\hat{a}_{\mathrm{ini}}-\int_{\mathcal{T}}\hat{b}_{\mathrm{ini}}(t)Z_n(t)\,dt\Big\}^\top\,.$$
 Subsequently, we can repeat the first step with $Y_i$ replaced by $Y_i-\hat{\theta}_{\mathrm{ini}}^\top X_i$ to update the estimators of $\hat{a}_{\mathrm{ini}}$ and $\hat{b}_{\mathrm{ini}}$. This procedure is iterated until the outcome regression (OR) estimators of $a$, $b$ and $\theta$, denoted by $\hat{a}_{\mathrm{OR}}$, $\hat{b}_{\mathrm{OR}}$ and $\hat{\theta}_{\mathrm{OR}}$, are stabilized. More specifically, let $\hat{a}_{\rm{OR}}^{(t)},\hat{b}_{\rm{OR}}^{(t)}$ and $\hat{\theta}_{\rm{OR}}^{(t)}$ denote respectively the estimators of $a,b$ and $\theta$ at the $t$-th iteration. We set our final estimators as $\hat{a}_{\rm{OR}}=\hat{a}_{\rm{OR}}^{(t+1)},\hat{b}_{\rm{OR}}=\hat{b}_{\rm{OR}}^{(t+1)}$ and $\hat{\theta}_{\rm{OR}}=\hat{\theta}_{\rm{OR}}^{(t+1)}$ if $\{|\hat{a}_{\rm{OR}}^{(t+1)}-\hat{a}_{\rm{OR}}^{(t)}|^2+\|\hat{b}_{\rm{OR}}^{(t+1)}-\hat{b}_{\rm{OR}}^{(t)}\|_{L^2}^2+ \|\hat{\theta}_{\rm{OR}}^{(t+1)}-\hat{\theta}_{\rm{OR}}^{(t)}\|_{\mathbb{R}^p}^2\}^{1/2}/\{|\hat{a}_{\rm{OR}}^{(t)}|^2+\|\hat{b}_{\rm{OR}}^{(t)}\|_{L^2}^2+ \|\hat{\theta}_{\rm{OR}}^{(t)}\|_{\mathbb{R}^p}^2\}^{1/2}<0.05$ or $t$ reaches 100 (a prespecfied maximum number of iterations). Here, $\|\cdot\|_{L^2}$ and $\|\cdot\|_{\mathbb{R}^p}$ denote the $L^2$ norm in $L^2(\mathcal{T})$ and the Euclidean norm in $\mathbb{R}^p$, respectively.  

The OR estimation is easy to implement but requires strong parametric assumptions, while the FSW estimator requires fewer modelling assumptions but is subject to a slow convergence rate. Thus, it is desirable to develop an estimator that possesses more attractive properties by combining these two.
Note that \eqref{id:DR} satisfies the so-called doubly robust (DR) property: for two generic functions $\tilde{E}(Y|X,Z)$ and $\tilde{\pi}(X,Z)$, it holds that
\begin{align}\label{eqDR}
E\{Y^*(z)\} = E[\{Y-\tilde{E}(Y|X,Z)\} \tilde{\pi}(X,Z)+E_{X}\{ \tilde{E}(Y|X,Z)\} | Z=z]\,,
\end{align}
if either $\tilde{E}(Y|X,Z)=E(Y|X,Z)$ or $\tilde{\pi}(X,Z)=\pi_0(X,Z)$ but not necessarily both are satisfied; see Appendix~\ref{AppscProofDR} for the derivation.

Under model~\eqref{eq1}, \eqref{eqDR} equals $a+\int_\mathcal{T} b(t) z(t)\, dt$. To estimate $a$ and $b$, it suffices to conduct the functional linear regression as in Section~\ref{sc32} to regress an estimator of $\{Y-E(Y|X,Z)\} \pi(X,Z)+E_{X}\{ E(Y|X,Z)\}$ on $Z$. Specifically, we estimate $E(Y|X,Z)$ by the OR estimator, $\hat{E}(Y|X,Z)=\hat{a}_{\mathrm{OR}}+\int_\mathcal{T} \hat{b}_{\mathrm{OR}}(t) Z(t)\,dt +\hat{\theta}_{\mathrm{OR}}^\top X$, and $\pi_0(X,Z)$ by $\hat{\pi}_{hk}(X,Z)$. We define the DR estimator of $b$ as $
\hat{b}_{\mathrm{DR}}(t) = \sum_{j=1}^q \hat{b}_{\mathrm{DR},j} \hat{\phi}_j(t)$, where $\hat{b}_{\mathrm{DR},j} = \hat{\lambda}_j^{-1} \hat{e}_{\mathrm{DR},j}$ with $\hat{e}_{\mathrm{DR},j}=\int_\mathcal{T} \hat{e}_{\mathrm{DR}} (t) \hat{\phi}_j(t)\,dt$. The expression of $\hat{e}_{\mathrm{DR}}(t)$ is 
\begin{align*}
\dfrac{1}{n}\sum_{i=1}^n \bigg[ \{ Y_i-\hat{E}(Y| X_i,Z_i)\} \hat{\pi}_{hk}(X_i,Z_i)+\dfrac{1}{n}\sum_{j=1}^n\{ \hat{E} (Y| X_j,Z_i)\} - \hat{\mu}_{Y,\mathrm{DR}}\bigg]\{Z_i(t)-\hat{\mu}_Z(t)\}\,,
\end{align*}
where
\begin{align*}
\hat{\mu}_{Y,\mathrm{DR}} = \dfrac{1}{n}\sum_{i=1}^n \bigg[ \{ Y_i-\hat{E} (Y| X_i,Z_i)\} \hat{\pi}_{hk}(X_i,Z_i)+\dfrac{1}{n}\sum_{j=1}^n\{ \hat{E} (Y| X_j,Z_i)\}\bigg]\,.
\end{align*}
The DR estimator of $a$ is defined as $\hat{a}_{\mathrm{DR}}=\hat{\mu}_{Y,\mathrm{DR}}-\int_\mathcal{T} \hat{b}_{\mathrm{DR}} \hat{\mu}_Z(t)\,dt$.

Provided that model~\eqref{eq1} is correctly specified, according to the DR property~\eqref{eqDR}, the DR estimator $\hat{E}_{\mathrm{DR}}\{Y^*(z)\}=\hat{a}_{\mathrm{DR}}+\int_\mathcal{T} \hat{b}_{\mathrm{DR}}(t) z(t)\,dt$ is consistent if either $\hat{E}(Y|X,Z)$ or $\hat{\pi}_{hk}(X,Z)$ is consistent. The consistency of $\hat{E}(Y|X,Z)$ mainly depends on the correct specification of the partially linear model~\eqref{MixedLinear}, while the consistency of $\hat{\pi}_{hk}(X,Z)$, as a nonparametric estimator, relies on less restrictive (but more technical) assumptions.  

\section{Asymptotic Properties}\label{sc4}
We investigate the asymptotic convergence rates of the proposed three estimators, while assuming that $Z$ is fully observed.
%and the case of sparsely observed $Z$ is left for future research.
Under the main model is $E\{Y^*(z)\}= a + \int_\mathcal{T} b(t) z(t)\,dt$ in~\eqref{eq1}, although the estimators of $a$ are based on those of $b$, they are essentially empirical means, which can achieve the ${n}^{-1/2}$ convergence rate \citep{Shin2009}. Thus, our primary focus is to derive the convergence rates of our estimators of $b$, which, depending on the smoothness of $Z$ and $b$, are slower than those in the finite-dimensional settings.

Let $\mathcal{X}\subset \mathbb{R}^p$ denote the support of $X$ and $\mathcal{Z}\subset L^2(\mathcal{T})$ the support of $Z$. To derive the convergence rate of $\hat{b}_{\mathrm{FSW}}$, we first provide the uniform convergence rate of the estimator $\hat{\pi}_{hk}$ of the FSW $\pi_0$ over $\mathcal{X}$ and $\mathcal{Z}$, which requires the following conditions.

\paragraph{Condition A}

\begin{enumerate}

\item[\namedlabel{CB1}{(A1)}] The set $\mathcal{X}$ is compact. For all $z\in \mathcal{Z}$ and $x\in \mathcal{X}$, $\pi_0(z,x)$ is strictly bounded away from zero and infinity.

\item[\namedlabel{CB2}{(A2)}] For all $z\in \mathcal{Z}$, $\exists~\eta_z\in \mathbb{R}^k$ and a constant $\alpha>0$ such that $\sup_{(z,x)\in \mathcal{Z}\times\mathcal{X}}|\rho'^{(-1)}\{\pi_0(z,x)\}-\eta_z^\top \nu_k(x)|=O(k^{-\alpha})$, where $\rho'^{(-1)}$ is the inverse function of $\rho'$.

\item[\namedlabel{CB3}{(A3)}] The eigenvalues of $E\{ \nu_k(X)\nu_k(X)^\top  | Z=z\}$  are bounded away from zero and infinity uniformly in $k$ and $z$. There exists a sequence of constants $\zeta(k)$ satisfying $\sup_{x\in \mathcal{X}}\|\nu_k(x) \|\leq \zeta(k)$.

\item[\namedlabel{CB4}{(A4)}] There exists a continuously differentiable function $\mu(h)>0$, for $h>0$, such that for all $z\in\mathcal{Z}$, $P\{d(Z,z)\leq h\}/\mu(h)$ is strictly bounded away from zero and infinity. There exists $h_0>0$ such that $\sup_{h\in(0,h_0)}\mu'(h)$ is bounded.

\item[\namedlabel{CB5}{(A5)}] The kernel $K:\mathbb{R}\to \mathbb{R}^+$ is bounded, Lipschitz and supported on $[0,1]$ with $K(1)>0$.

%\item[\namedlabel{CB6}{(A6)}] For $g_{k_1,k_2,z}(X)=\rho'\{\eta_{z}^\top \nu_k(X)\}v_{k_1}(X)$ and $v_{k_1}(X)v_{k_2}(X)$, there exists $\lambda>0$ such that {\color{red}for any fixed $k$} and all $k_1,k_2 = 1,\ldots,k$, all $z_1,z_2\in\mathcal{Z}$ and all $m\geq 2$, $\big| E[g_{k_1,k_2,z_1}(X)|Z=z_1]-E[g_{k_1,k_2,z_2}(X)|Z=z_2]\big| / d^{\lambda}(z_1,z_2)$ and $E[| g_{k_1,k_2,z_1}(X) |^m|Z=z_1]$ are bounded.

\item[\namedlabel{CB6}{(A6)}] For $g_{k_1,k_2,z}(X)=\rho'\{\eta_{z}^\top \nu_k(X)\}v_{k_1}(X)$ and $v_{k_1}(X)v_{k_2}(X)$ with $k_1,k_2=1,\ldots,k$, there exists $\lambda>0$ such that for all  $m\geq 2$, $\big| E[g_{k_1,k_2,z_1}(X)|Z=z_1]-E[g_{k_1,k_2,z_2}(X)|Z=z_2]\big| / d^{\lambda}(z_1,z_2)$ and $E[| g_{k_1,k_2,z_1}(X) |^m|Z=z_1]$ are uniformly bounded over $k,z_1$ and $z_2$.

\item[\namedlabel{CB7}{(A7)}] Let $\psi_{\mathcal{Z}}(\epsilon)$ be the Kolmogorov $\epsilon$-entropy of $\mathcal{Z}$, i.e.,~$\psi_{\mathcal{Z}}(\epsilon)=\log\{N_\epsilon(\mathcal{Z})\}$, where $N_\epsilon(\mathcal{Z})$ is the minimal number of open balls in $L^2(\mathcal{T})$ of radius $\epsilon$ (with the semi-metric $d$) covering $\mathcal{Z}$. It satisfies $\sum_{n=1}^\infty \exp[ -(\delta-1) \psi_{\mathcal{Z}}\{(\log n) /n\}]<\infty$ for some $\delta>1$, and $(\log n)^2/\{n \mu(h)\} < \psi_{\mathcal{Z}}\{(\log n) /n\}<n\mu(h)/\log n$ for $n$ large enough.
\end{enumerate}

Condition~\ref{CB1} requires that $X$ is compactly supported, which can be relaxed by restricting the tail behaviour of the distribution of $X$ at the cost of more tedious technical arguments. The boundedness of $\pi_0$ in Condition~\ref{CB1} (or some equivalent condition) is also commonly required in the literature \citep{Kennedy2017, Ai2021, d2021overlap}. This condition can be relaxed if other smoothness assumptions are made on the potential outcome distribution \citep{ma2020robust}. Condition~\ref{CB2} essentially assumes that the convergence rate of the sieve approximation $\eta_z^\top \nu_k(x)$ for  $\rho'^{(-1)}\{\pi_0(z,x)\}$ is polynomial. This can be satisfied, for example, with $\alpha=+\infty$ if $X$ is discrete, and with $\alpha=s/r$ if $X$ has $r$ continuous components and $\nu_k$ is a power series, where $s$ is the degree of smoothness of $\rho'^{(-1)}\{\pi_0(z,\cdot)\}$ with respect to the continuous components in $X$ for all $z\in L^2(\mathcal{T})$ \citep{Chen2007}. Further, Condition~\ref{CB3} ensures that the sieve approximation conditional on a functional variable is not degenerate. Similar conditions are routinely assumed in the literature of sieve approximation~\citep{Newey1997}. Conditions~\ref{CB4} to \ref{CB6} are standard in functional nonparametric regression \citep{Ferraty2006,Ferraty2010}. In particular, the function $\mu(h)$ in Condition~\ref{CB4}, also referred to as the small ball probability, has been studied extensively in the literature~\citep{Li2001,Ferraty2006};  Condition~\ref{CB5} requires that $K$ is compactly supported, which is not satisfied for the Gaussian kernel but convenient for technical arguments; the variable $g_{k_1,k_2,z}(X)$ in Condition~\ref{CB6} serves as the role of the response variable in the traditional regression setting. Condition~\ref{CB7} is less standard: it regularizes the support $\mathcal{Z}$ by controlling its Kolmogorov entropy, which is used to establish the uniform convergence rate on $\mathcal{Z}$. This condition is satisfied, for example, if $\mathcal{Z}$ is a compact set, $d(\cdot,\cdot)$ is a projection metric and $(\log n)^2=O\{n \mu(h)\}$; see \citet{Ferraty2010} for other examples.

\begin{theo}\label{theo2}
Under Conditions~\ref{CB1} to \ref{CB7}, assuming that $k\to \infty$ and $h\to 0$ as $n\to \infty$, we have
\begin{align*}
\sup_{(z,x)\in\mathcal{Z}\times \mathcal{X}}| \hat{\pi}_{hk}(z,x)-\pi_0(z,x) |=O\bigg(\zeta(k)\bigg[ k^{-\alpha}+h^\lambda \sqrt{k} +  \sqrt{\dfrac{\psi_\mathcal{Z}\{(\log n)/n\}k}{n\mu(h)}} \bigg] \bigg)
\end{align*}
holds almost surely.
\end{theo}

Theorem~\ref{theo2} provides the sup-norm convergence rate of $\hat{\pi}_{hk}$ over the support of $X$ and $Z$. The proof is given in Appendix~\ref{Appsc3}. The term $\zeta(k)(k^{-\alpha}+h^\lambda \sqrt{k})$ and the term $\zeta(k) \sqrt{\psi_\mathcal{Z}\{(\log n)/n\} k/\{n\mu(h)\}}$ can be viewed as the convergence rates of the bias and standard deviation, respectively.

To provide the convergence rate of $\hat{b}_{\mathrm{FSW}}$, we recall that $E\{\pi_0(Z,X)Y|Z=z\}=a+\int_\mathcal{T} b(t) z(t)\,dt$ under model~\eqref{eq1}. Let $\epsilon_{\pi}=\pi_0(Z,X)Y-a-\int_\mathcal{T} b(t) Z(t)\,dt$ be the residual variable.  Let $\gamma>1$ and $\beta>\gamma/2+1$ be two constants and the $c_j$'s be some generic positive constants.

\paragraph{Condition B}

\begin{enumerate}
\item[\namedlabel{CC1}{(B1)}] The variable $\epsilon_\pi$ has a finite fourth moment, i.e.,~$E(\epsilon_\pi^4 )< \infty$.

\item[\namedlabel{CA1}{(B2)}] The functional variable $Z$ has a finite fourth moment, i.e.,~$\int_{\mathcal{T}}E\{Z^4(t)\}\,dt< \infty$. The PC score $\xi_j$ defined in \eqref{eq:KLexpansion} satisfies $E(\xi_j^4)\leq c_1 \lambda_j^2$, and the eigenvalue $\lambda_j$ defined in \eqref{eq:EigenDecomposition} satisfies $\lambda_j-\lambda_{j+1}\geq c_2^{-1} j^{-\gamma-1}$ for all $j\geq 1$.

\item[\namedlabel{CA2}{(B3)}] The coefficient $b_j$ satisfies $|b_j|\leq c_3j^{-\beta}$, for all $j\geq 1$.
\end{enumerate}

Condition~\ref{CC1} makes a mild restriction on the moment of $\epsilon_\pi$.
Condition~\ref{CA1} imposes regularity conditions on the random process $Z$, which requires that the differences of the adjacent eigenvalues do not decay too fast. Condition~\ref{CA2} assumes an upper bound for the decay rate of $b_j$, the coefficients of $b$  projected on $\{\phi_j\}_{j=1}^\infty$. The latter two conditions are adopted from~\citet{Hall2007} for technical purposes.

\begin{theo}\label{theo3}
Under Conditions \ref{CB1} to \ref{CB7}, \ref{CC1} to \ref{CA2} and model~\eqref{eq1}, assuming that $k\to \infty$ and $h\to 0$ as $n \to \infty$ and choosing  the truncation parameter $q\asymp n^{1/(\gamma+2\beta)}$, we have
\begin{align*}
\int_\mathcal{T} \{\hat{b}_{\mathrm{FSW}}(t)-b(t)\}^2\,dt = O_p \bigg( \bigg[ k^{-2\alpha}+h^{2\lambda} k + \dfrac{\psi_\mathcal{Z}\{(\log n)/n\}k}{n\mu(h)} \bigg] \zeta^2(k) n^{(\gamma+1)/(\gamma+2\beta)}\bigg).
\end{align*}
\end{theo}

The proof of Theorem~\ref{theo3} is given in Appendix~\ref{Appsc4}. The convergence rate of $\int_\mathcal{T} \{\hat{b}_{\mathrm{FSW}}(t)-b(t)\}^2\,dt$ in Theorem~\ref{theo3} can be quite slow mainly because of the small ball probability function $\mu(h)$. In the case of Gaussian process endowed with a metric, $\mu(h)$ has an exponential decay rate of $h$ \citep{Li2001}, which leads to a non-infinitesimal rate of $\int_\mathcal{T} \{\hat{b}_{\mathrm{FSW}}(t)-b(t)\}^2\,dt$. However, $\mu(h)$ can be much larger by choosing a proper semi-metric $d$ \citep{Ferraty2006,Ling2018}, so that the final convergence rate of $\int_\mathcal{T} \{\hat{b}_{\mathrm{FSW}}(t)-b(t)\}^2\,dt$ reaches a polynomial decay of $n$. Another way to improve the convergence rate is to impose additional structure assumptions. For example, if we assume that the slope function $b$ can be fully characterized by a finite number of PC basis functions, or the random process $Z$ essentially lies on a finite dimensional (not necessarily Euclidean) space \citep{Lin2021}, then the truncation parameter $q$ does not need to tend to infinity and the convergence rate of $\int_\mathcal{T} \{\hat{b}_{\mathrm{FSW}}(t)-b(t)\}^2\,dt$ can be much faster. Without such assumptions, estimating $b$ in an infinite dimensional space with a nonparametric estimator of $\pi_0$ leads to a cumbersome convergence rate.

The OR estimator $\hat{b}_{\mathrm{OR}}$ relies on functional principal component analysis and the partially linear model~\eqref{MixedLinear}. Additional conditions are needed to derive the convergence rate of $\hat{b}_{\mathrm{OR}}$.

\paragraph{Condition C}

\begin{enumerate}
\item[\namedlabel{CA3}{(C1)}] The covariate $X$ has finite fourth moment, i.e.,~$E(\|X\|^4)<\infty$. For $m=1,\ldots,p$ and $j\geq 1$, $|\Cov \{X_m,\int_\mathcal{T} Z(t) \phi_j(t) \,dt\}|<c_4 j^{-\gamma-\beta}$.

\item[\namedlabel{CA4}{(C2)}] Let $g_m(t)=\sum_{j=1}^\infty \Cov \{X_m,\int_\mathcal{T}  Z(t) \phi_j(t)\,dt\} \phi_j(t)/\lambda_j$ and $u_{im} = X_{im}-E(X_m)-\int_\mathcal{T} g_m(t) \{Z_i(t)-\mu_Z(t)\}\,dt$. For $m=1,\ldots,p$, $E(u_{1m}^2|Z_1)< \infty$, and $E(\mathfrak{u}_1^\top \mathfrak{u}_1)$ is positive definite with $\mathfrak{u}_1=(u_{11},\ldots,u_{1p})^\top$.
\end{enumerate}

Conditions~\ref{CA3} and \ref{CA4} are adopted from~\citet{Shin2009}, which make mild assumptions on covariate $X$ to ensure $\sqrt{n}$-consistency of the estimated coefficient $\hat{\theta}_{\mathrm{OR}}$.

\begin{theo} \label{theo1}
Under Conditions \ref{CA1}, \ref{CA2}, \ref{CA3}, \ref{CA4} and model~\eqref{MixedLinear}, choosing  the truncation parameter $q\asymp n^{1/(\gamma+2\beta)}$, we have
\begin{align*}
\int_\mathcal{T} \{\hat{b}_{\mathrm{OR}}(t)-b(t)\}^2\,dt = O_p(n^{(-2\beta+1)/(\gamma+2\beta)})\,.
\end{align*}
\end{theo}

Because $\hat{b}_{\mathrm{OR}}$ is theoretically equivalent to the least square estimator proposed by~\citet{Shin2009} provided that the estimators of each component in the additive model are unique~\citep{Buja1989}, the theorem above is a direct result from Theorem 3.2 in~\citet{Shin2009} and thus its proof is omitted. The convergence rate of $\hat{b}_{\mathrm{OR}}$ is the same as that of the estimated slope function in the traditional functional linear regression \citep{Hall2007}, i.e., such rate remains unchanged despite of the additional estimation of $\theta$. Compared with Theorem \ref{theo3}, the convergence rate here $O_p(n^{(-2\beta+1)/(\gamma+2\beta)})=O_p(n^{-1})\cdot O_p(n^{(\gamma+1)/(\gamma+2\beta)})$ is much faster. Specifically, the part of the convergence rate $O_p \big(\zeta^2(k) \big[ k^{-2\alpha}+h^{2\lambda} k +  \psi_\mathcal{Z}\{(\log n)/n\}k / \{n\mu(h)\} \big] \big)$, which is the convergence rate of $\sup_{(z,x)\in\mathcal{Z}\times \mathcal{X}}| \hat{\pi}_{hk}(z,x)-\pi_0(z,x) |^2$, is replaced by a faster one, $O_p(n^{-1})$, because the nonparametric estimator $\hat{\pi}_{hk}$ is not used. 
%We can see from Theorem~\ref{theo2} that this rate is the convergence rate of $\sup_{(z,x)\in\mathcal{Z}\times \mathcal{X}}| \hat{\pi}_{hk}(z,x)-\pi_0(z,x) |^2$.

According to the discussion at the end of Section~\ref{sc33}, under model~\eqref{eq1} and the assumptions of Theorem~\ref{theo2}, $\hat{b}_{\mathrm{DR}}$ is consistent. If assumptions of Theorem~\ref{theo3} hold, it has the same convergence rate as $\hat{b}_{\mathrm{FSW}}$. If the stronger model assumption~\eqref{MixedLinear} holds, then $\hat{b}_{\mathrm{DR}}$ enjoys the same convergence rate as $\hat{b}_{\mathrm{OR}}$, provided that Condition C is satisfied. This differs from classical doubly robust estimators, whose consistency relies on correctly specifying one of the two parametric models. 

\section{Selection of Tuning Parameters}\label{sccom}

For the FSW estimator, our methodology requires a kernel function $K$ with a metric $d$, basis functions $\nu_k$ and tuning parameters $h$, $k$ and $q$. Using a compactly supported kernel, $K$ would make the denominator of the first term in \eqref{eq22.5} too small for some $Z_i$ and destabilize the maximization of $H_{hk,Z_i}$ in \eqref{eq22.5}. Therefore, we suggest using the Gaussian kernel, $K(u)=\exp(-u^2)$, with the semi-metric $d$ being the $L^2$ norm. To develop a general estimation strategy for $\pi_0$, %that is applicable for a generic dataset, 
we standardize the $X_i$'s and use a unified set of choices for $\nu_k$. Letting $X_{\mathrm{min}} = (\min_i \{X_{i1}\}_{i=1}^n,\ldots,\min_i \{X_{ip}\}_{i=1}^n)^\top$, $X_{\mathrm{max}} = (\max_i \{X_{i1}\}_{i=1}^n,\ldots,\max_i \{X_{ip}\}_{i=1}^n)^\top$, we transform the $X_i$'s as follows,
\begin{align*}
X_{i,\mathrm{st}}=\dfrac{2(X_i-X_{\mathrm{min}})}{X_{\mathrm{max}}-X_{\mathrm{min}}}-1\,,
\end{align*}
so that they have support on $[-1,1]$. Let $X_{ij,\mathrm{st}}$ be the $j$-th component of $X_{i,\mathrm{st}}$ and $\mathcal{P}_\ell$ be the $\ell$-th standard Legendre polynomial on $[-1,1]$. If $k$ is given, we define $v_{1}(X_i)=1$ and $v_{p(\ell-1)+j+1}(X_i)=\mathcal{P}_{\ell}(X_{ij,\mathrm{st}})$, for $i=1,\ldots,n$, $j=1,\ldots,p$ and $\ell = 1,\ldots,(k-1)/p$. We restrict the range of $k$ such that $(k-1)/p$ is a positive integer, and orthonormalize the resulting matrix $\{\nu_{k}(X_1) ,\ldots,\nu_{k}(X_n) \}^\top$.

One may select $h,k$ and $q$ jointly by an $L$-fold cross-validation (CV), but this may be computationally intensive. We suggest a two-stage CV procedure as follows. In the first stage, we select the truncation parameter $q$ using the OR estimator. Specifically, we randomly split the dataset into $L$ parts, $\mathcal{S}_1,\ldots,\mathcal{S}_L$. Let $\mathcal{S}_{-\ell}$ denote the remaining sample with $\mathcal{S}_\ell$ excluded. We define the CV loss as
\begin{align*}
\mathrm{CV}_{L}^{\textrm{OR}}(q) = \sum_{\ell=1}^L \sum_{i\in \mathcal{S}_\ell} \bigg[Y_i - \hat{a}_{-\ell,\textrm{OR}} - \sum_{j=1}^{q} \hat{b}_{j,-\ell,\textrm{OR}}\bigg\{ \int_\mathcal{T}\hat{\phi}_j(t) \hat{\mu}_Z(t)\,dt+ \hat{\xi}_{ij} \bigg\} -\hat{\theta}^\top_{-\ell,\textrm{OR}}X_i \bigg]^2 \,,
\end{align*}
where $\hat{a}_{-\ell,\textrm{OR}}$, $\hat{b}_{j,-\ell,\textrm{OR}}$ and $\hat{\theta}_{-\ell,\textrm{OR}}$ are obtained by applying the method in Section~\ref{sc33} to $\mathcal{S}_{-\ell}$. We choose the truncation parameter as $\hat{q}_{\mathrm{CV}}= \argmin_{q\in\{1,\ldots,q_{99}\}}\mathrm{CV}_{L}^{\textrm{OR}}(q)$, where $q_{99}$ is the number of principal components that can explain at least 99\% of the variance of $Z$.

Noting that the rate requirements for $q$ in Theorem~\ref{theo2} and Theorem~\ref{theo3} are the same, we thus use the same $\hat{q}_{\mathrm{CV}}$ in the FSW estimator. It remains to select $h$ and $k$ in the second stage. We consider
\begin{align*}
&\mathrm{CV}_L^{\mathrm{FSW}}(h,k) \\
&= \sum_{\ell=1}^L \sum_{i\in \mathcal{S}_\ell} \bigg[Y_i\hat{\pi}_{hk,-\ell}(Z_i,X_i)- \hat{a}_{-\ell,\mathrm{FSW}} - \sum_{j=1}^{\hat{q}_{\mathrm{CV}}} \hat{b}_{j,-\ell,\mathrm{FSW}}\bigg\{ \int_\mathcal{T}\hat{\phi}_j(t) \hat{\mu}_Z(t)\,dt+ \hat{\xi}_{ij} \bigg\} \bigg]^2 \,,
\end{align*}
where $\hat{\xi}_{ij} = \int_\mathcal{T} \{Z_i(t)-\hat{\mu}_Z(t)\}\hat{\phi}_j(t)\,dt$ is the estimated PC score, $\hat{a}_{-\ell,\mathrm{FSW}}$, $\hat{b}_{j,-\ell,\mathrm{FSW}}$ and $\hat{\pi}_{hk,-\ell}$ are obtained by applying the method in Section~\ref{sc32} to $\mathcal{S}_{-\ell}$. We choose the one from a candidate set of $\{h,k\}$ such that $\mathrm{CV}_L(h,k)$ is minimized.

As for the DR estimator, we still need to select the truncation parameter $q$. This can be achieved by minimizing the CV loss,
\begin{align*}
\mathrm{CV}_L^{\mathrm{DR}}(q) &= \sum_{\ell=1}^L \sum_{i\in \mathcal{S}_\ell} \bigg[\{ Y_i-\hat{E}_{-\ell}(Y| X_i,Z_i)\} \hat{\pi}_{hk,-\ell}(X_i,Z_i)+\dfrac{1}{|\mathcal{S}_{\ell}|}\sum_{j\in \mathcal{S}_{\ell}} \{ \hat{E}_{-\ell}(Y| X_j,Z_i)\}\\
&\quad - \hat{a}_{-\ell,\mathrm{DR}} - \sum_{j=1}^{q} \hat{b}_{j,-\ell,\mathrm{DR}}\bigg\{ \int_\mathcal{T}\hat{\phi}_j(t) \hat{\mu}_Z(t)\,dt+ \hat{\xi}_{ij} \bigg\} \bigg]^2 \,,
\end{align*}
where $|\{\cdot\}|$ denotes the number of elements in the set $\{\cdot\}$, $\hat{a}_{-\ell,\mathrm{DR}}$, $\hat{E}_{-\ell}(Y| X_i,Z_i)$, $\hat{\pi}_{hk,-\ell}$ and $\hat{b}_{j,-\ell,\mathrm{DR}}$ are obtained by applying the methods in Section~\ref{sc33} to $\mathcal{S}_{-\ell}$.

\section{Simulation Study}\label{scSim}
In addition to the methods of functional stabilized weight (FSW), outcome regression (OR) and double robustness (DR) developed in Section~\ref{sc3}, we also consider the method using the nonparametric principal component weight (PCW) proposed by~\citet{Zhang2021}, which assumes the same functional linear model for the ADRF;  the weight-modified kernel ridge regression (WMKRR), a nonlinear method proposed by~\citet{Wang2023}; as well as the direct functional linear regression (FLR) and kernel ridge regression (KRR) of $Y$ on $Z$. The FLR and KRR are expected to be biased if there are confounding effects. We select the tuning parameters $h,k$ and $q$ following the procedure in Section~\ref{sccom}, and choose the number of PCs used in estimating the PCW so as to explain 95\% of the variance of $Z$ following \citet{Zhang2021}. For a fair comparison, we use the truncation parameter $\hat{q}_{\mathrm{CV}}$ in Section~\ref{sccom} for estimating $b$ in all methods that utilize the functional linear model. The code to reproduce the simulation results for our methods and the FLR is available at \url{https://github.com/ruoxut/FunctionalTreatment}.

\begin{table}[t]
\caption{Mean (standard deviation) of the empirical mean squared error (MSE) of the ADRF with the best values highlighted in boldface.}\label{tb1}
\centering
\resizebox{\columnwidth}{!}{
%\begin{threeparttable}
\begin{tabular}{ccccccc}
\hline
   $n=200$  & (\romannumeral1) & (\romannumeral2) &(\romannumeral3) &(\romannumeral4) &(\romannumeral5) &(\romannumeral6)   \\
\hline 
FSW &1.14 (0.66) & \bf{2.15 (1.36)} & 8.40 (8.60) & \bf{9.69 (10.08)} &  \bf{7.94 (12.86)}  & 180.24 (42.74)\\
OR & 1.11 (0.68) & 6.96 (3.42) & 6.85 (9.77) & 11.55 (14.11) &  29.52 (11.79)  & 192.79 (58.15)\\
DR & \bf{1.08 (0.66)} & 7.56 (3.65) & \bf{6.65 (9.56)} & 11.34 (14.36) & 29.27 (12.12)  & 186.10 (52.82)\\
PCW & 2.68 (1.90) & 12.98 (2.93) & 13.10 (7.74) & 16.13 (8.31) & 16.17 (9.72)  & 186.84 (46.58)\\
FLR & 4.93 (1.63) & 9.78 (2.29) & 21.53 (7.90) &22.98 (7.72) & 37.98 (14.08)   & 205.45 (61.24) \\
WMKRR & 8.30 (3.23) & 22.67 (4.72) & 25.07 (10.54) & 25.57 (10.67) &  
 34.71 (14.18)    & \bf{133.79 (40.30)}\\
KRR & 6.41 (1.83) & 11.84 (2.42) &  25.94 (6.59) & 27.63 (6.60) & 79.56 (21.87) & 343.84 (158.34) \\
\hline
\end{tabular}
}
\resizebox{\columnwidth}{!}{
\begin{tabular}{ccccccc}
\hline
   $n=500$  & (\romannumeral1) & (\romannumeral2) &(\romannumeral3) &(\romannumeral4) &(\romannumeral5) &(\romannumeral6)   \\
\hline 
FSW & 0.50 (0.34) & \bf{1.36 (0.95)} & 5.82 (4.27) & \bf{4.57 (2.98)} &  \bf{3.60 (14.88)}  & 173.47 (28.29) \\
OR & 0.42 (0.33) & 6.42 (2.58) & 1.61 (2.52) & 5.66 (2.53) &  25.33 (6.13)  & 169.20 (41.27) \\
DR & \bf{0.38 (0.31)} & 6.86 (2.74) & \bf{1.52 (2.45)} & 5.64 (2.41) &  25.39 (6.55)  & 166.22 (36.65) \\
PCW & 1.68 (1.08) & 12.16 (1.85)& 8.42 (3.54) & 11.50 (3.05) &  12.09 (7.21)  & 169.64 (34.34) \\
FLR & 4.31 (1.05) & 9.00 (1.51) & 17.74 (3.95) & 18.83 (3.60) &  33.14 (8.37)& 174.16 (46.45)\\
WMKRR & 6.62 (2.37) & 19.62 (3.52) & 23.89 (9.77) & 24.80 (10.01) & 35.26 (11.86)   & \bf{125.12 (32.51)}\\
KRR & 5.14 (1.06) & 11.10 (1.54) & 25.06 (4.22) & 26.38 (4.33) & 79.08 (14.62)  & 302.76 (79.42) \\
\hline
\end{tabular}
}
%\begin{tablenotes}
%\small
%\item \hspace{-0.9cm} FSW: functional stabilized weight; OR: outcome regression; DR: double robustness; PCW: principal component weight.
%\end{tablenotes}
%\end{threeparttable}
\end{table}

We consider the sample sizes $n=200$ and $500$, and generate data for models (\romannumeral1) to (\romannumeral5) as follows. For $i=1,\ldots,n$, we generate the functional treatment $Z_i(t)$, for $t\in[0,1]$, by $Z_i(t) = \sum_{j=1}^6 A_{ij}\phi_j(t)$, where $A_{i1}=4U_{i1}$, $A_{i2}=2\sqrt{3}U_{i2}$, $A_{i3} = 2\sqrt{2}U_{i3}$, $A_{i4}=2U_{i4}$, $A_{i5}=U_{i5}$, $A_{i6}=U_{i6}/\sqrt{2}$ with the $U_{ij}$'s being the independent standard normal variables, $\phi_{2m-1}(t)=\sqrt{2} \sin(2m\pi t)$ and $\phi_{2m}(t)=\sqrt{2} \cos(2m\pi t)$ for $m=1,2,3$. For the slope function $b$, we define $b(t) = 2 \phi_1(t)+ \phi_2(t)+ \phi_3(t)/2 + \phi_4(t)/2$. The $Z_i$'s and $b$ are the same as those in~\citet{Zhang2021}. We consider five models for the covariate $X_i$ and the outcome variable $Y_i$:
\begin{enumerate}
\item[(\romannumeral1)] $X_i=U_{i1}+\epsilon_{i1}$ and $Y_i=1+\int_0^1 b(t)Z_i(t)\,dt+2X_i+\epsilon_{i2}$,

\item[(\romannumeral2)] $X_i=U_{i1}+0.25\epsilon_{i1}$ and $Y_i=1+\int_0^1 b(t)Z_i(t)\,dt+5 \sin (X_i)+\epsilon_{i2}$,

\item[(\romannumeral3)] $X_i=(X_{i1},X_{i2})^\top =\{(U_{i1}+1)^2+\epsilon_{i1},U_{i2} \}^\top$ and $Y_i=1+\int_0^1 b(t)Z_i(t)\,dt+2  X_{i1}+2 X_{i2}+\epsilon_{i2}$,

\item[(\romannumeral4)] $X_i=(X_{i1},X_{i2})^\top =\{(U_{i1}+1)^2+\epsilon_{i1},U_{i2} \}^\top$ and $Y_i=1+\int_0^1 b(t)Z_i(t)\,dt+2 X_{i1}+2 \cos(X_{i1})+5.5\sin(X_{i2})+\epsilon_{i2}$,
 
\item[(\romannumeral5)]  $X_i=(X_{i1},X_{i2})^\top =\{(U_{i1}+1)^2+\epsilon_{i1},U_{i2} \}^\top$ and $Y_i=1+\int_0^1 b(t)Z_i(t)\,dt+\{\int_0^1 b(t)Z_i(t)\,dt\}^2/25+2 X_{i1}+5.5\sin(X_{i2})+\epsilon_{i2}$,
\end{enumerate}
where $\epsilon_{i1}\sim N(0,1)$ and $\epsilon_{i2}\sim N(0,25)$ are generated independently.  In addition, we also consider a nonlinear model from~\citet{Wang2023},
\begin{enumerate}
    \item[(\romannumeral6)] $X_i=(U_{i11},U_{i21},U_{i31},U_{i41})^\top$, $Y_i=10\{U_{i21}U_{i11}^2+U_{i41}^2 \sin(2U_{i31})\}+0.5A_{i1}^2+4\sin(A_{i1})$, $Z_i(t) = \sum_{j=1}^4 A_{ij} \sqrt{2}\sin(2j\pi t)$, for $t\in[0,1]$, where $A_{i1} = 4U_{i11}+U_{i12}$, $A_{i2} = 2\sqrt{3} U_{i21}+U_{i22}$, $A_{i3} = 2\sqrt{2} U_{i31}+U_{i32}$, $A_{i4} = 2U_{i41}+U_{i42}$, with the $U_{ijk}$'s being the independent standard normal variables.
\end{enumerate}

The PC scores $A_{ij}$ of $Z_i(t)$ affect  
covariate $X_i$ linearly in models (\romannumeral1) and (\romannumeral2) and nonlinearly in models (\romannumeral3) and (\romannumeral4), and 
covariate $X_i$ affects the outcome variable $Y_i$ linearly in models (\romannumeral1) and (\romannumeral3) and nonlinearly in models (\romannumeral2) and  (\romannumeral4). The functional linear model~\eqref{eq1} for the ADRF is correctly specified for models (\romannumeral1) to (\romannumeral4), while the partially linear model~\eqref{MixedLinear} is only correctly specified for models (\romannumeral1) and (\romannumeral3). Under models (\romannumeral5) and (\romannumeral6), the functional linear model~\eqref{eq1} for the ADRF is misspecified. Due to the confounding effect of $X$, the FLR and KRR ignoring $X$ are expected to be biased. For each combination of model and sample size, we replicate 200 simulations and evaluate the results by  the empirical mean squared error (MSE) of the ADRF following \citet{Wang2023}, $\mathrm{MSE} = \sum_{i=1}^n [\hat{E} \{Y^*(Z_i)\} - E \{Y^*(Z_i)\}]^2 /n$, where $\hat{E}(\cdot)$ denotes a generic estimator of the ADRF. 
 
%\begin{figure}[t]
%\begin{center}
%\hspace*{-.2cm}
%\includegraphics[width=4.5cm]{(1)n200_box_full.pdf}\hspace*{-.4cm}
%\includegraphics[width=4.5cm]{(2)n200_box_full.pdf}\hspace*{-.4cm}
%\includegraphics[width=4.5cm]{(3)n200_box_full.pdf}\hspace*{-.4cm}
%\includegraphics[width=4.5cm]{(4)n200_box_full.pdf}\hspace*{-.4cm}\\
%\hspace*{-0.2cm}
%\includegraphics[width=4.5cm]{(1)n500_box_full.pdf}\hspace*{-.4cm}
%\includegraphics[width=4.5cm]{(2)n500_box_full.pdf}\hspace*{-.4cm}
%\includegraphics[width=4.5cm]{(3)n500_box_full.pdf}\hspace*{-.4cm}
%\includegraphics[width=4.5cm]{(4)n500_box_full.pdf}
%\caption{Box plots of the ISEs of all methods under models (\romannumeral1) to (\romannumeral4) from left to right and for $n=200$ (top) and $n=500$ (bottom).}\label{fg001}
%\end{center}
%\end{figure}

Table~\ref{tb1} summarizes the mean and the standard deviation of the  MSEs for all configurations. 
Our proposed estimators FSW, OR and DR outperform the other methods under models (\romannumeral1)--(\romannumeral4). In particular, our three estimators are close to each other and are better than the others under models (\romannumeral1) and (\romannumeral3), where covariate $X$ affects the outcome variable $Y$ linearly. Under models (\romannumeral2) and (\romannumeral4) where the partially linear model~\eqref{MixedLinear} is misspecified, the FSW performs the best, followed by the OR and DR, and it is worth noting that the latter two still perform better than the other methods. Under model (\romannumeral5) where the functional linear model is misspecified, the FSW remarkably performs the best, followed by the PCW, while the FLR and WMKRR perform similarly. Under model (\romannumeral5), the FSW is expected to be biased. The performance of PCW is better than FLR but inferior to our methods under models (\romannumeral1)--(\romannumeral4). The WMRKK performs the best under model (\romannumeral6), but its performance is not satisfactory under models (\romannumeral1)--(\romannumeral5).

\begin{figure}[t]
  \begin{center}  
  \includegraphics[width=.33\textwidth]{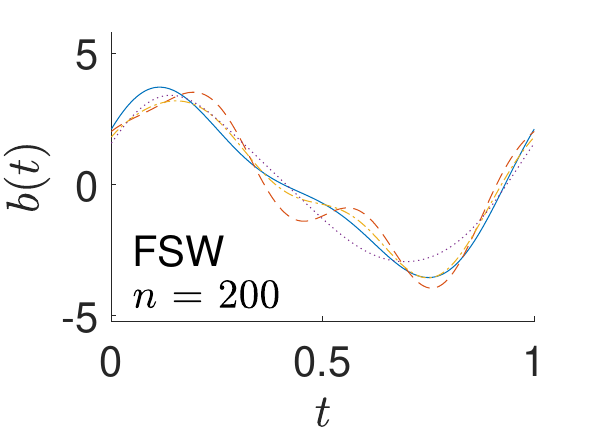}\hspace*{-.35cm}
  \includegraphics[width=.33\textwidth]{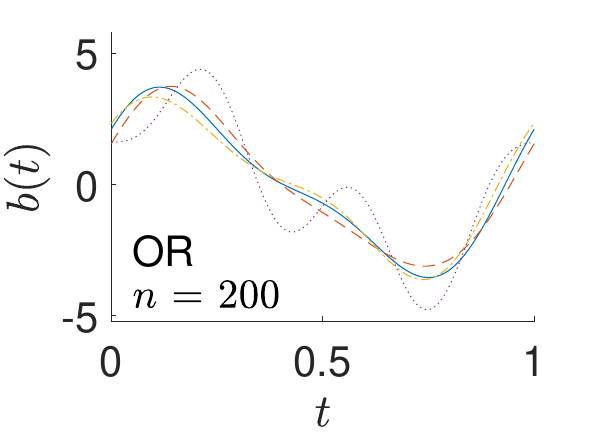}\hspace*{-.35cm}
  \includegraphics[width=.33\textwidth]{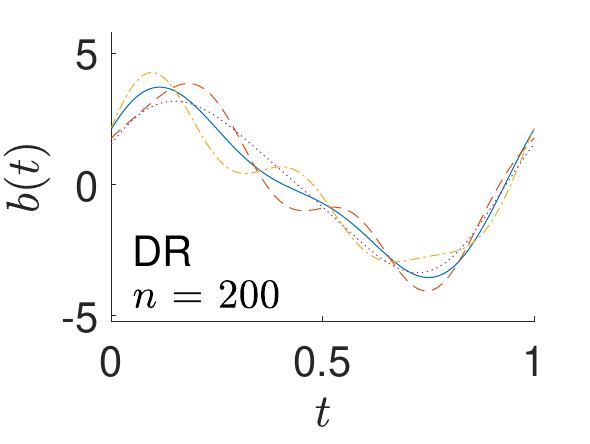}
  \\
  \includegraphics[width=.33\textwidth]{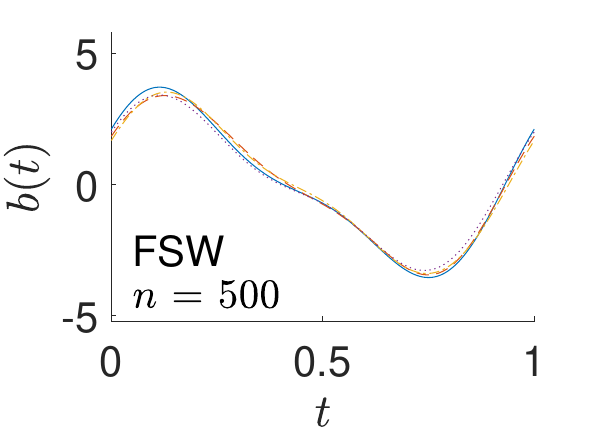}\hspace*{-.35cm}
  \includegraphics[width=.33\textwidth]{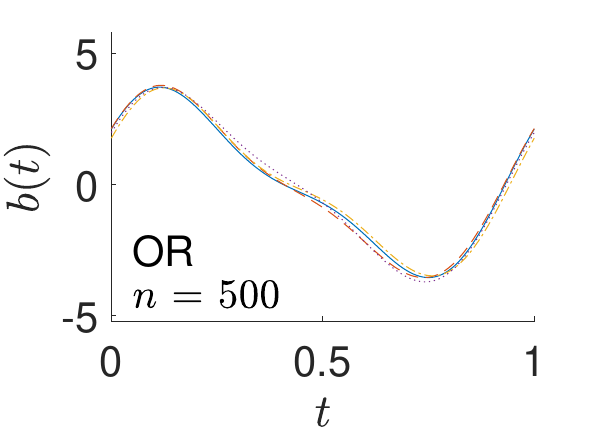}\hspace*{-.35cm}
  \includegraphics[width=.33\textwidth]{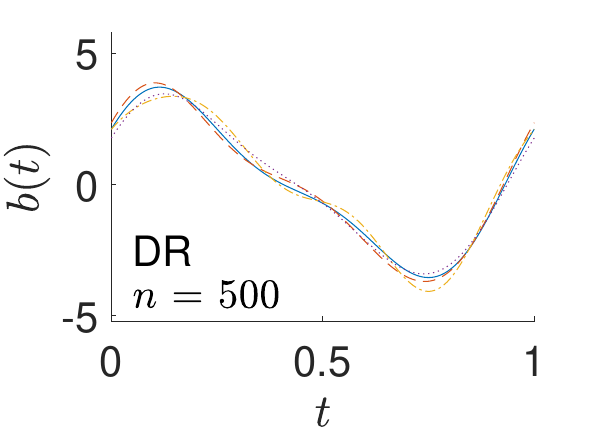}
  \\
  \includegraphics[width=.33\textwidth]{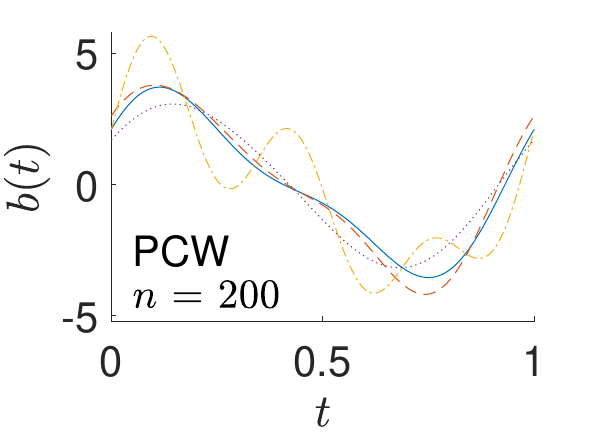}
  \includegraphics[width=.33\textwidth]{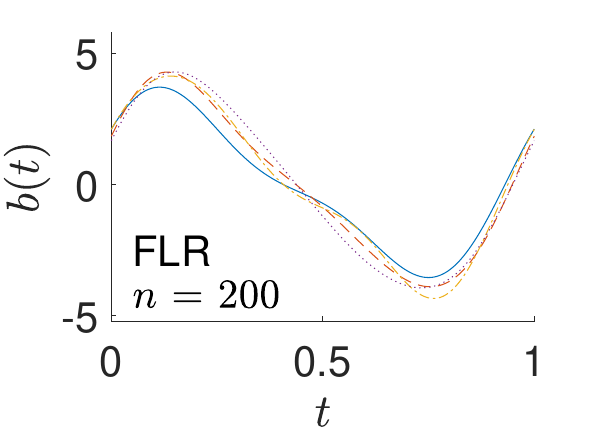} 
  \\
  \includegraphics[width=.33\textwidth]{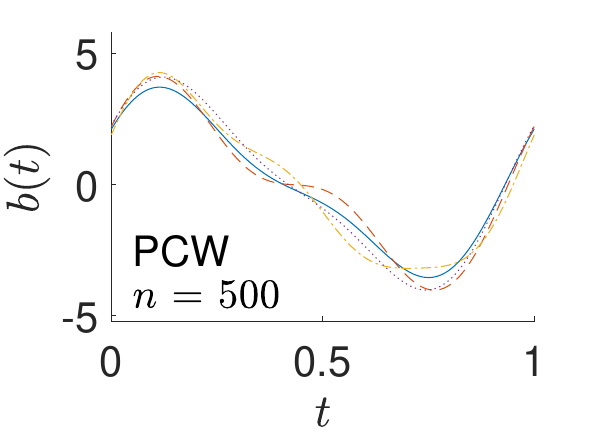}
  \includegraphics[width=.33\textwidth]{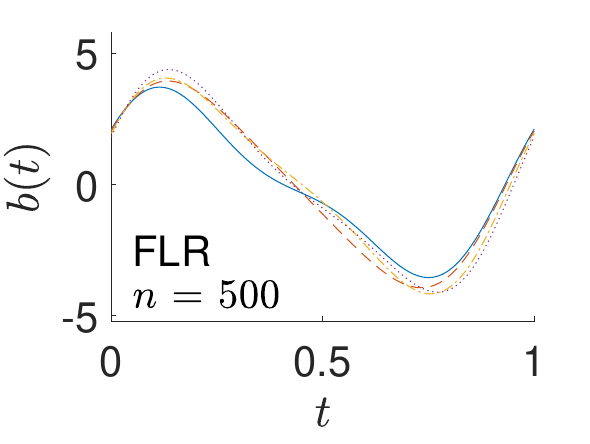} 
  \caption{True curve (\----), first (\---~\---~\---), second (\---\,$\cdot$\,\---) and third ($\cdot$~$\cdot$~$\cdot$) quartile estimated slope functions under model (i) with $n=200$ and 500 for all methods.}\label{fg01}
  \end{center}
  \end{figure} 

\begin{figure}[t]
\begin{center} 
\includegraphics[width=.33\textwidth]{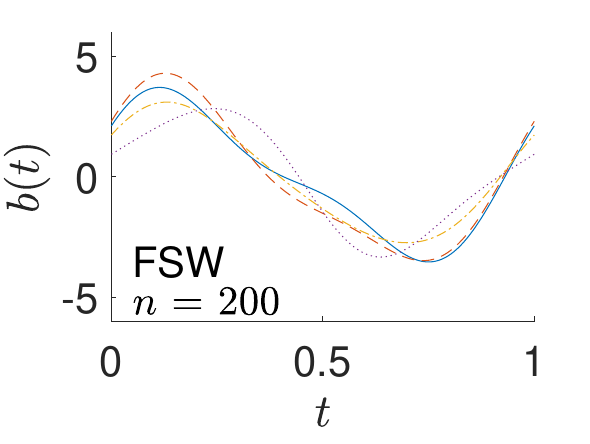}\hspace*{-.35cm}
  \includegraphics[width=.33\textwidth]{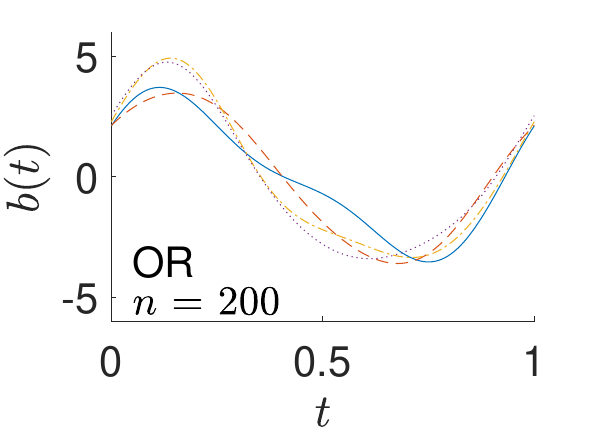}\hspace*{-.35cm}
  \includegraphics[width=.33\textwidth]{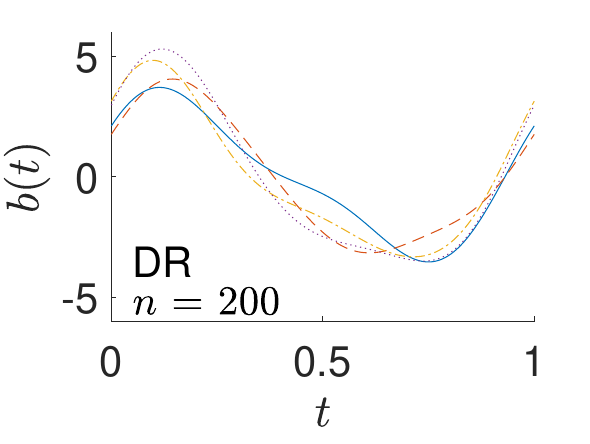}
  \\
  \includegraphics[width=.33\textwidth]{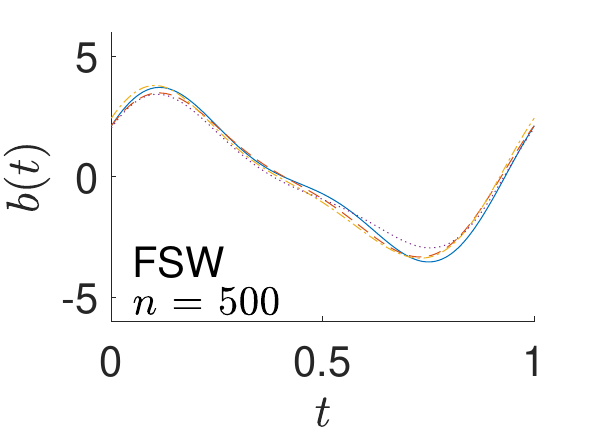}\hspace*{-.35cm}
  \includegraphics[width=.33\textwidth]{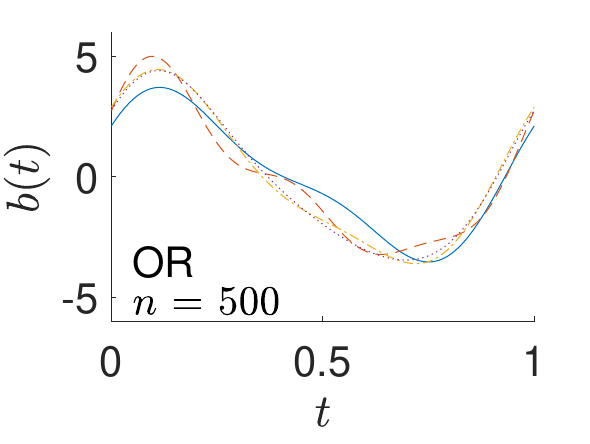}\hspace*{-.35cm}
  \includegraphics[width=.33\textwidth]{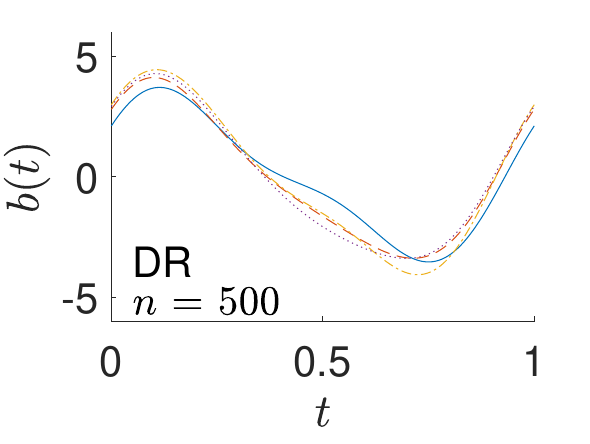}
  \\
  \includegraphics[width=.33\textwidth]{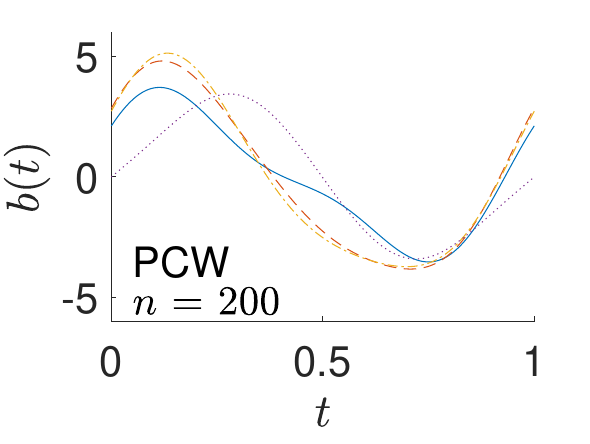}
  \includegraphics[width=.33\textwidth]{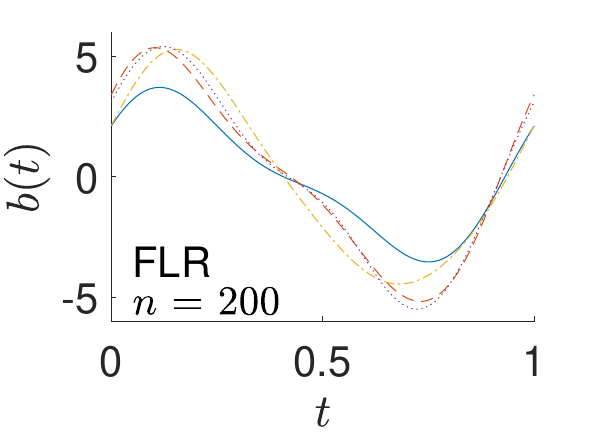} 
  \\
  \includegraphics[width=.33\textwidth]{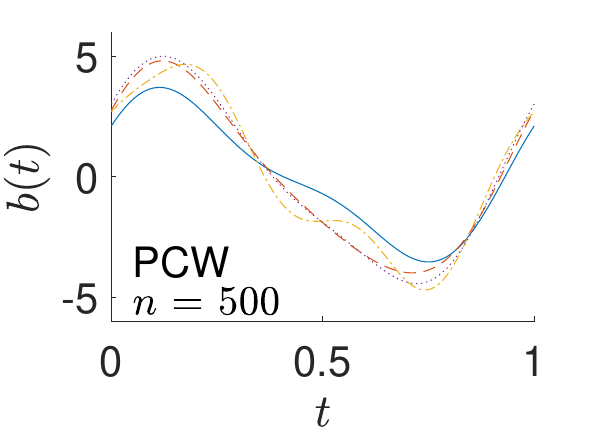}
  \includegraphics[width=.33\textwidth]{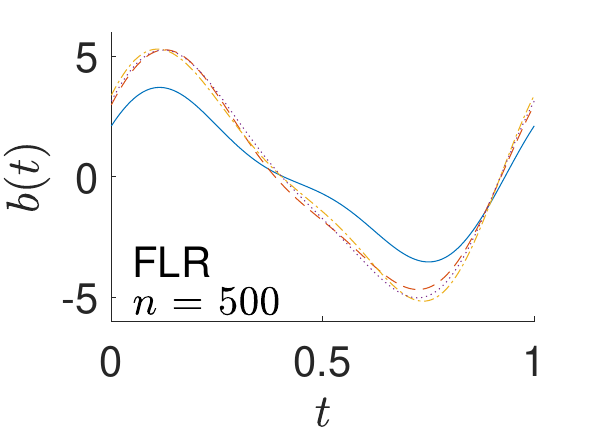} 
\caption{True curve (\----), first (\---~\---~\---), second (\---\,$\cdot$\,\---) and third ($\cdot$~$\cdot$~$\cdot$) quartile estimated slope functions under model (iv) with $n=200$ and 500 for all methods.}\label{fg02}
\end{center}
\end{figure}

To give visualization of representative estimated slope functions, we show the quartile curves of the samples corresponding to the first, second and third quartile values of 200 MSEs for model (i) in Figure~\ref{fg01}. It can be seen that the estimated curves using the FSW, OR and DR are close to the truth when $n=500$, while the PCW is more variable and the FLR is more biased. Figure~\ref{fg02} corresponds to the results under model (iv), i.e., a more challenging model structure. The differences among the estimators are more significant: FSW performs the best, followed by the DR, OR and PCW, while the FLR is more biased than the others.

We also conduct a small-scale running time comparison of our estimator FSW and the WMKRR proposed by \citet{Wang2023}. Both estimators contain the same weight function $\pi_0$ estimated by different nonparametric methods. We compare the running time of computing the two estimators of $\pi_0$, including selection of the tuning parameters. The experiments were carried out on a PC with an i7-12700 CPU and 16 GB RAM. As an illustration, model (\romannumeral1) is used as the data generation model with $n=100,500$ and $1000$. We repeat the experiments 10 times. The values of averaged running time in seconds are 6.19 ($n=100$), 37.28 ($n=500$) and 236.88 ($n=1000$) for the FSW, while the corresponding values for the WMKRR are 36.27, 204.77 and 440.67. We see that, for the given model, computing the estimator of $\pi_0$ using the FSW is much faster than that using the WMKRR.

\section{Real Data Analysis}\label{scReal}

\begin{figure}[t]
  \begin{center}
  \hspace*{-.4cm}
  \includegraphics[width=0.35\textwidth]{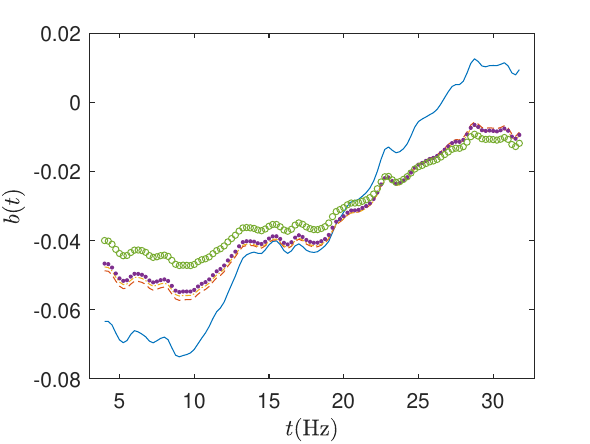}\hspace*{-.4cm}
  \includegraphics[width=0.35\textwidth]{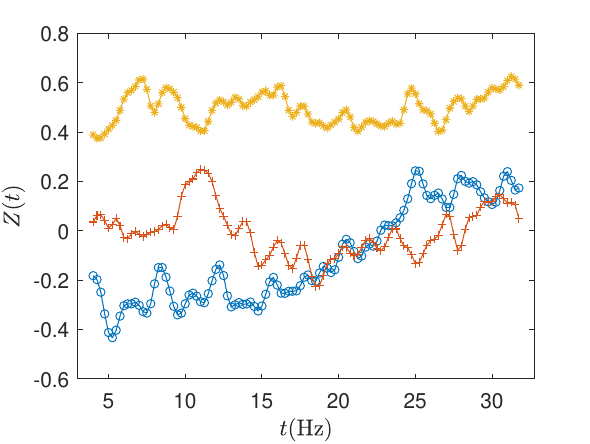}\hspace*{-.4cm}
  \includegraphics[width=0.35\textwidth]{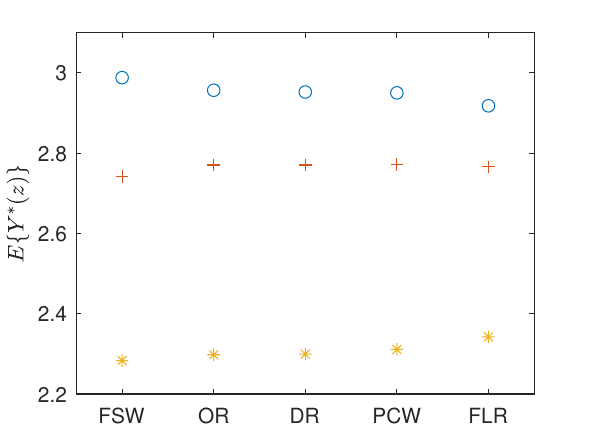}
  \caption{Left: the estimated slope functions using FSW (---), OR (-- --), DR (-- $\cdot$ --), PCW ($\cdots$) and FLR ($\circ$); middle: the frontal asymmetry curves corresponding to the smallest ($\circ$), median ($+$) and largest ($*$) average values over 4 to 15 Hz; right: the estimated $E\{Y^*(Z)\}$ of the three selected curves with marks corresponding to the middle panel.}\label{fg2}
  \end{center}
\end{figure}

We illustrate the estimation of the ADRF using the five methods, FSW, OR, DR, PCW and FLR, on the electroencephalography (EEG) dataset from~\citet{Ciarleglio2022}. The EEG is a relatively low-cost tool to measure human neuronal activities. The measurements are taken from multiple electrodes placed on the scalps of subjects, and they are then processed and transformed to produce current source density curves on the frequency domain, which can  provide information on the intensity of neuronal activities. In particular, the frontal asymmetry curves, considered potential biomarkers for major depressive disorder, are treated as the functional treatment variable $Z(t)$, for $t\in [4,31.75]$ Hz. The outcome variable $Y$ is defined as $\log(\tilde{Y}+1)$, where $\tilde{Y}$ is the quick inventory of depressive symptomatology (QIDS) score measuring the severity of depressive symptomatology. A larger value of $Y$ indicates a more severe depressive disorder. We investigate the causal effect of neuronal activities represented by the frontal asymmetry curves on the severity of major depressive disorder. Potential confounding covariates $X=(X_1,X_2,X_3)^\top$ include age $X_1$, sex $X_2$ ($1$ for female and $0$ for male) and Edinburgh Handedness Inventory score $X_3$ (ranging from $-100$ to $100$ corresponding to completely left to right handedness). Individuals with missing $Z$ (29.6\% of the total sample) are removed, which results in a sample of size 85 males and 151 females. The means (standard deviations) of $Y$, $X_1$ and $X_3$ are 2.73 (0.72), 35.97 (13.07) and 72.69 (48.76), respectively. For comparison of the estimated ADRF using different methods, the confounding variables $X$ are centralized.

We use the number of PCs explaining 95\% of the variance of $Z$ (equal to $11$ for this dataset) in estimating the PCW. We choose the truncation parameter $q$ by $\hat{q}_{\mathrm{CV}}$ to estimate the slope function $b$ for all methods and choose $h$ and $k$ used in FSW by minimizing $\mathrm{CV}_{L}^{\mathrm{FSW}}$ as in Section~\ref{sccom}. As a result, $\hat{q}_{\mathrm{CV}}=2$ is used, which can explain about 80\% of the variance of $Z$.
%The first two principal component basis functions are plotted in Figure~\ref{fg3}.

%\begin{figure}[t]
%\begin{center}
%\includegraphics[width=8cm]{EEG_phi_1.pdf}
%\includegraphics[width=8cm]{EEG_phi_2.pdf}
%\caption{The first (left) and second (right) principal components basis functions of $Z$.}\label{fg3}
%\end{center}
%\end{figure}

The left panel of Figure~\ref{fg2} shows the estimated slope functions. All the estimated slope functions have the same increasing trend over frequencies, while they have a negative effect on the outcome variable for low frequencies. In other words, subjects with higher frontal asymmetry values on the low-frequency domain tend to be healthier. 
This is consistent with the observations in~\citet{Ciarleglio2022}, where it was found a negative association between the frontal asymmetry curves and the major depressive disorder status using their functional regression model. More specifically, in the left panel of Figure~\ref{fg2}, the estimated slope function of FSW is the steepest and shows the largest effect in the low-frequency domain. In contrast, the estimated slope function of FLR has the overall smallest absolute values; the other three curves are quite close to each other.

To visualize the above negative effect in the estimated slope function, we compute the estimated ADRF $E\{Y^*(z)\}$ of three curves, which correspond to the smallest, median and largest average frontal asymmetry values over low frequencies 4 to 15 Hz. The middle panel of Figure~\ref{fg2} exhibits the selected curves and the right panel shows the corresponding estimated $E\{Y^*(z)\}$ using all methods, which are around 2.9, 2.8 and 2.3 for the three curves, respectively.

\section{Discussion}\label{sc7}
In the case of a functional treatment variable, we propose three estimators of the ADRF, namely, the FSW, OR and DR estimators, based on the functional linear model for the ADRF. The consistency of the FSW estimator relies on the functional linear model of $E\{Y^*(z)\}$ by developing a nonparametric estimator of the weight; while the OR estimator requires a more restrictive linear model for $Y$ regressed on $(Z,X)$. The DR estimator is consistent if either of the first two estimators is consistent. 
%Asymptotic convergence rates are provided, and numerical results demonstrate the great advantages of our methods.

It is of interest to construct the confidence band for the slope function to better quantify the ADRF or ATE, which, however, is difficult even in the simpler context of functional linear regression. As shown in \citet{Cardot2007}, it is impossible for an estimator of the slope function to converge in distribution to a nondegenerate random element under the norm topology. For the OR estimator, it is possible to adapt the method proposed by \citet{Imaizumi2019} to our context to construct an approximate confidence band. For the other two estimators, the construction of the confidence band is challenging due to the less restrictive modelling assumption and the nonparametric estimator of the FSW, which warrants further investigation.

\section*{Acknowledgements}
We thank the Action	Editor and two referees for their careful reviews that greatly improved our work. We also thank Dr.~Jiayi Wang for sharing her code. The research of Zheng Zhang is
supported by the funds from the National Key R\&D Program of China [grant number 2022YFA1008300] and the National Natural Science Foundation of China [grant number 12371284]. Tan's research was supported by NSFC (No.~12401363 and 12471263), the Fundamental Research Funds for the Central Universities, and Key Laboratory of Intelligent Computing and Applications (Ministry of Education).
Yin's research was supported by the Research Grants Council of Hong Kong (17308321), the
Theme-based Research Scheme (TRS) Project T45-401/22-N, and the Patrick SC Poon Endowment Fund.
\newpage

\appendix

\section{Proof of Proposition~\ref{prop1} }\label{Appsc2}

\begin{proof}
The `if' part follows from~\eqref{eq2}, and thus we show the `only if' part below. Suppose
\begin{align*}
E\{\pi(z,X)v(X)|Z=z\}=E\{v(X)\}\,,
\end{align*}
for all integrable functions $v$. Since $E\{v(X)\}=E\{\pi_0(z,X)v(X)|Z=z\}$ for all integrable functions $v$, we have
\begin{align*}
E[\{\pi(z,X)-\pi_0(z,X)\} v(X)|Z=z\}]=0\,.
\end{align*}
Taking $v(X)=\exp(i a^\top X)$ for all $a\in \mathbb{R}^p$, we have
\begin{align*}
E[\{\pi(z,X)-\pi_0(z,X)\} \exp(ia^\top X)|Z=z\}]=0\,,
\end{align*}
According to the uniqueness of the Fourier transformation, we conclude that $\pi(z,\cdot)=\pi_0(z,\cdot)$ a.s.

\end{proof}

\section{Generalized empirical likelihood interpretation of $\widehat{\pi}_{hk}$}\label{Appsc1}
To gain more insight on the function $\rho$, we investigate the generalized empirical likelihood interpretation of our estimator. We  show that the estimator defined by~\eqref{eq22.5} is the dual solution to the following local generalized empirical likelihood maximization problem. For each $Z_i$, $i=1,\ldots,n$,
\begin{align}\label{eq24}
\begin{aligned}
&\max_{\{\pi(Z_i,X_j)\}_{j\in \mathcal{S}_{-i}}} - \dfrac{\sum_{j\in \mathcal{S}_{-i} } D\{\pi(Z_i,X_j)\}K\{d(Z_j,Z_i) / h\}}{\sum_{j\in \mathcal{S}_{-i}}  K\{d(Z_j,Z_i) / h\} } \,,\\
&\mathrm{s.t.}~\dfrac{\sum_{j\in \mathcal{S}_{-i}}  \pi(Z_i,X_j)\nu_k(X_j)K\{d(Z_j,Z_i) / h\} }{\sum_{j\in \mathcal{S}_{-i}}  K\{d(Z_j,Z_i) / h\}} = \dfrac{1}{n-1} \sum_{j\in \mathcal{S}_{-i}}  \nu_k(X_j)\,,
\end{aligned}
\end{align}
where $D(w)$ is a distance measure between $w$ and 1 for $w\in\mathbb{R}$. The function $D(v)$ is continuously differentiable satisfying $D(1)=0$ and  $\rho(-u) = D\{ D^{'(-1)}(u)\} - u\cdot D^{'(-1)}(u)$, where $D^{'(-1)}$ is the inverse function of $D^{'}$, the first derivative of $D$. Different choices of $\rho$ correspond to different distance measures. For example, if we take $\rho(v) = -\exp(-v-1)$, then $D(v) = -v \log(v)$ is the information entropy and the weights correspond to the exponential tilting~\citep{Imbens1998}. Other choices include
$\rho(v) = \log(v)+1$ with $D(v)=-\log(v)$, the empirical likelihood~\citep{Owen1988}, and $\rho(v) = -(1 - v)^2/2$ with $D(v)=(1-v)^2/2$, yielding the implied weights of the continuous updating of the generalized method of moments~\citep{Hansen1996}.

Following~\citet{Tseng1991}, we show that the dual problem of~\eqref{eq24} is to maximize $H_{hk,Z_i}$ in~\eqref{eq22.5}. For each $i=1,\ldots,n$, we let $\bar{\nu}_k=\sum_{j\in \mathcal{S}_{-i}}\nu_k(X_j)/(n-1)$, $K_{ij} = K\{d(Z_j,Z_i)/h\}/[\sum_{j\in \mathcal{S}_{-i}}K\{d(Z_j,Z_i)/h\}]$, $\pi_{ij}=\pi(Z_i,X_j)$ and $w_{ij}=K_{ij}\pi_{ij}$, for $j\in \mathcal{S}_{-i}$. Moreover, let $D_{ij}(\nu)=D(\nu/K_{ij})$, $V_k=\big(\nu_k(X_j)\big)_{j\in \mathcal{S}_{-i}}\in \mathbb{R}^{k\times(n-1)}$, ${w}_i=(w_{ij})_{j\in \mathcal{S}_{-i}}^\top$, and $F({w}_i)=\sum_{j\in \mathcal{S}_{-i}}  D_{ij}(w_{ij})K_{ij}$. Then,~\eqref{eq24} can be written as
\begin{align}\label{eqa1}
\begin{aligned}
\min_{{w}_i} F({w}_i)\,,{\textrm{~s.t.~}} V_k {w}_i=\bar{\nu}_k\,.
\end{aligned}
\end{align}

We define the conjugate convex function of $F$ as
\begin{align*}
F^*({u})&=\sup_{{w}_i} \sum_{j\in \mathcal{S}_{-i}}\{u_j w_{ij}-K_{ij}D_{ij}(w_{ij})\}\\
&=\sup_{\{\pi_{ij}\}_{j\in \mathcal{S}_{-i}}}  \sum_{j\in \mathcal{S}_{-i}} K_{ij} \{u_j\pi_{ij} - D(\pi_{ij})\}\\
&=\sum_{j\in \mathcal{S}_{-i}} K_{ij} \{u_j\pi^*_{ij} - D(\pi^*_{ij})\}\,,
\end{align*}
where the $\pi_{ij}^*$'s satisfy
\begin{align*}
u_j = D^{'}(\pi^*_{ij}) \Rightarrow \pi^*_{ij} = D^{'(-1)}(u_j), \ j = 1,\ldots, n\,,
\end{align*}
by taking the first order condition. It follows that
\begin{align}\label{eqa2}
F^*({u})=-\sum_{j\in \mathcal{S}_{-i}}K_{ij}\rho(-u_j)\,,
\end{align}
where $\rho(-u) = D\{D^{'(-1)}(u)\} - u\cdot D^{'(-1)}(u)$.

Following~\citet{Tseng1991} and using~\eqref{eqa2}, we conclude that the dual problem of~\eqref{eq24} is
\begin{align*}
\max_{\eta\in \mathbb{R}^k}\{-F^*(\eta^\top V_k)+\eta^\top \bar{\nu}_k \}&=\max_{\eta\in \mathbb{R}^k}\sum_{j\in \mathcal{S}_{-i}}\bigg[ K_{ij}\rho\{-\eta^\top \nu_k(X_j)\}+\dfrac{\eta^\top \nu_k(X_j)}{n-1}\bigg]\\
&=\max_{\eta\in \mathbb{R}^k}\sum_{j\in \mathcal{S}_{-i}}\bigg[ K_{ij}\rho\{\eta^\top \nu_k(X_j)\}-\dfrac{\eta^\top \nu_k(X_j)}{n-1}\bigg]=\max_{\eta\in \mathbb{R}^k}H_{hk,Z_i}(\eta)\,.
\end{align*}

\section{Proof of Double Robustness in~\eqref{eqDR}}\label{AppscProofDR}
Recall from Section~\ref{sc2} that $E\{Y^*(z)\}=E\{Y\pi_0(X,Z)|Z=z\}=E\{E(Y|X,Z=z)\}$.
First assuming that $\tilde{E}(Y|X,Z) = E(Y|X,Z)$, then we have
\begin{align*}
&E[\{Y-\tilde{E}(Y|X,Z)\} \tilde{\pi}(X,Z)+E_{X}\{ \tilde{E}(Y|X,Z)\} | Z=z]\\
=\,& E[\{Y- E (Y|X,Z)\} \tilde{\pi}(X,Z)|Z=z] + E\{E(Y|X,Z=z)\}\\
=\,& E\{Y^*(z)\}\,,
\end{align*}
where the second equality follows from the law of total expectation.

Assuming that $\tilde{\pi}(X,Z) = \pi_0(X,Z)=f_X(X) / f_{X|Z}(X|Z)$, we have
\begin{align*}
&E[\{Y-\tilde{E}(Y|X,Z)\} \tilde{\pi}(X,Z)+E_{X}\{ \tilde{E}(Y|X,Z)\} | Z=z]\\
=\,&E\{Y\pi_0(X,Z)|Z=z\}-E\{\tilde{E}(Y|X,Z)\pi_0(X,Z)|Z=z \}+E_{X}\{ \tilde{E}(Y|X,Z)\} | Z=z]\\
=\,&E\{Y^*(z)\}-\int \tilde{E}(Y|X=x,Z=z) f_{X}(x) \, d {x}+\int \tilde{E}(Y|X=x,Z=z) f_{X}(x) \, d {x}\\
=\,&E\{Y^*(z)\}\,.
\end{align*}
Therefore, the doubly robust property is proved.

\section{Proof of Theorem~\ref{theo2}}\label{Appsc3}

\begin{proof}
Recall from Section~\ref{sc32} that
\begin{align*}
\hat{\pi}_{hk}(Z_i,X_i) =  \rho'\{ \hat{\eta}_{Z_i}^\top \nu_{k}(X_i) \}\,,
\end{align*}
where $\hat{\eta}_{Z_i}$ is a $k$-dimensional vector maximizing the strictly concave function $H_{hk,Z_i}$ defined by
\begin{align*}
H_{hk,Z_i}(\eta) = \dfrac{\sum_{j\in \mathcal{S}_{-i}} \rho\{ \eta^\top \nu_{k}(X_j)\} K\{d(Z_j,Z_i)/h\} }{\sum_{j\in \mathcal{S}_{-i}} K\{d(Z_j,Z_i)/h\}} -\eta^\top\bigg\{ \dfrac{1}{n-1} \sum_{j\in \mathcal{S}_{-i}} \nu_k(X_j)\bigg\}\,.
\end{align*}
For a fixed $z\in L^2(\mathcal{T})$, we have
\begin{align*}
\hat{\pi}_{hk}(z,X_i) =  \rho'\{ \hat{\eta}_{z}^\top \nu_{k}(X_i) \}\,,
\end{align*}
where $\hat{\eta}_{z}$ is a $k$-dimensional vector maximizing $H_{hk,z}$ defined by
\begin{align}\label{eqb0.1}
H_{hk,z}(\eta) = \dfrac{\sum_{j\in \mathcal{S}_{-i}} \rho\{ \eta^\top \nu_{k}(X_j)\} K\{d(Z_j,z)/h\} }{\sum_{j\in \mathcal{S}_{-i}} K\{d(Z_j,z)/h\}} -\eta^\top\bigg\{ \dfrac{1}{n-1} \sum_{j\in \mathcal{S}_{-i}} \nu_k(X_j)\bigg\}\,.
\end{align}
Here, $\mathcal{S}_{-i}$ is independent of $Z_i$ for $i=1,\ldots,n$.

Let $H_{hk,z}^*(\eta)$ be the theoretical counterpart of $H_{hk,z}(\eta)$,
\begin{align*}
H_{hk,z}^*(\eta) = \dfrac{E[ \rho\{ \eta^\top \nu_{k}(X)\} K\{d(Z,z) / h\}] }{E\{ K\{d(Z,z)/ h\}\}} -\eta^\top E\{ \nu_k(X)\}\,,
\end{align*}
$\eta_z^*=\argmax_{\eta\in \mathbb{R}^p}H_{hk,z}^*(\eta)$ and  $ \pi_{hk}^*(z,x) =  \rho'\{ \eta_{z}^{*\top} \nu_{k}(x) \}$.

Note that
\begin{align*}
&\underset{(z,x)\in \mathcal{Z}\times \mathcal{X}}{\sup} |\hat{\pi}_{hk}(z,x)-\pi_0(z,x) |\\
\leq\,& \underset{(z,x)\in \mathcal{Z}\times \mathcal{X}}{\sup} |\hat{\pi}_{hk}(z,x)-\pi_{hk}^*(z,x) |+\underset{(z,x)\in \mathcal{Z}\times \mathcal{X}}{\sup} |\pi_{hk}^*(z,x)-\pi_0(z,x) |
\end{align*}
To prove Theorem~\ref{theo2}, it suffices to prove the convergence rates for the two terms on the right-hand side of the above inequality. Lemmas~\ref{lm2} and \ref{lm3} provide such results. Lemma~\ref{lm1} is needed to prove Lemmas~\ref{lm2} and \ref{lm3}.

\end{proof}

\begin{lem}\label{lm1}
Let
\begin{align*}
\hat{e}_{k_1,k_2}(z) =\dfrac{\sum_{j\in \mathcal{S}_{-i}} g_{k_1,k_2,z}(X_j) K\{d(Z_j,z)/ h\} }{(n-1)E[ K\{d(Z_j,z)/h\}]}\,,
\end{align*}
where $g_{k_1,k_2,z}$ is defined in \ref{CB6}, and
\begin{align*}
\hat{f}(z) =\dfrac{\sum_{j\in \mathcal{S}_{-i}} K\{d(Z_j,z) / h\} }{(n-1)E[ K\{d(Z_j,z)/h\}] }\,.
\end{align*}
Under Conditions~\ref{CB4} to \ref{CB7}, we have
\begin{align*}
&\underset{z\in \mathcal{Z} }{\sup}|\hat{f}(z)-1 |=O_{a.c.}\bigg\{\sqrt{\dfrac{\psi_{\mathcal{Z}}\{(\log n)/n\}}{n\mu(h)}}\bigg\}\,,\\
&\sum_{n=1}^\infty P\bigg(\inf_{z\in \mathcal{Z}}\hat{f}(z)<\dfrac{1}{2}\bigg) <\infty\,,\\
&\underset{z\in \mathcal{Z} }{\sup}|E\{\hat{e}_{k_1,k_2}(z)\}-E [g_{k_1,k_2,z}(X)|Z=z] |=O(h^\lambda)\,,\ \forall k_1,k_2\in  \{1,\ldots,k\},\\
&\underset{z\in \mathcal{Z} }{\sup}|\hat{e}_{k_1,k_2}(z)- E\{\hat{e}_{k_1,k_2}(z)\}|=O_{a.c.}\bigg\{\sqrt{\dfrac{\psi_{\mathcal{Z}}\{(\log n)/n\}}{n\mu(h)}}\bigg\}, \ \forall k_1,k_2\in  \{1,\ldots,k\},
\end{align*}
hold  uniformly  over $k $.
\end{lem}
\begin{proof}

Note that for all $k_1,k_2=1,\ldots,k$, $g_{k_1,k_2,z}$ is a scalar function, and Condition \ref{CB6} is assumed uniformly over $k$. Therefore, Lemmas 8 to 11 in \citet{Ferraty2010} can be used to derive the lemma above. Also, note the $O_{a.c.}$ denotes the almost complete convergence, which implies the almost sure convergence and convergence in probability.
\end{proof}

\begin{lem}\label{lm2}
Under Conditions~\ref{CB1} to \ref{CB6}, we have
\begin{align*}
\underset{(z,x)\in \mathcal{Z}\times \mathcal{X}}{\sup} |\pi_{hk}^*(z,x)-\pi_0(z,x)|=O\{\zeta(k)(k^{-\alpha}+h^\lambda \sqrt{k})\}
\end{align*}

\end{lem}

\begin{proof}
We first prove that
\begin{align}\label{eqb2}
\underset{(z,x)\in \mathcal{Z}\times \mathcal{X}}{\sup} |\pi_0(z,x)-\rho'\{\eta_z^\top \nu_k(x)\}| = O(k^{-\alpha})\,,
\end{align}
where $\eta_z$ is given in Condition~\ref{CB2}.

Recall that $\rho$ is a strictly increasing and concave function, so $\rho'^{(-1)}$ is strictly decreasing. Thus, we have
\begin{align*}
\rho'^{(-1)}(c_2) \leq \inf_{(z,x)\in \mathcal{Z}\times \mathcal{X}} \rho'^{(-1)}\{\pi_0(z,x)\}\leq \sup_{(z,x)\in \mathcal{Z}\times \mathcal{X}} \rho'^{(-1)}\{\pi_0(z,x)\} \leq \rho'^{(-1)}(c_1)\,,
\end{align*}
for some positive constants $c_1$, $c_2$, because $\pi_0(z,x)$ is bounded away from zero and infinity by Condition~\ref{CB1}. According to Condition~\ref{CB2}, there exists a constant $C_1>0$ such that
\begin{align*}
\sup_{(z,x)\in \mathcal{Z}\times \mathcal{X}}|\rho'^{(-1)}\{\pi_0(z,x)\}-\eta_z^\top \nu_k(x)|\leq C_1 k^{-\alpha}\,.
\end{align*}
Thus, we have, $\forall z \in \mathcal{Z}$ and $\forall x\in \mathcal{X}$,
\begin{align}\label{eqb2.5}
\begin{aligned}
\eta_z^\top \nu_k(x)&\in [\rho'^{(-1)}\{\pi_0(z,x)\}-C_1k^{-\alpha},\rho'^{(-1)}\{\pi_0(z,x)\}+C_1k^{-\alpha}] \\
&\subset [ \rho'^{(-1)}(c_2)-C_1k^{-\alpha},\rho'^{(-1)}(c_1)+C_1k^{-\alpha}]
\end{aligned}
\end{align}
and
\begin{align}\label{eqb3}
&\rho'\{\eta_z^\top \nu_k(x)+C_1 k^{-\alpha} \} - \rho'\{ \eta_z^\top \nu_k(x)\}\notag \\
<\,& \pi_0(z,x)- \rho'\{ \eta_z^\top \nu_k(x)\}\notag \\
<\,& \rho'\{\eta_z^\top \nu_k(x)-C_1 k^{-\alpha} \} - \rho'\{ \eta_z^\top \nu_k(x)\}\,.
\end{align}
By the mean value theorem, for $k$ large enough, there exist
\begin{align*}
\xi_1(z,x)\in (\eta_z^\top \nu_k(x),\eta_z^\top \nu_k(x)+C_1 k^{-\alpha})\subset [\rho'^{(-1)}(c_2)-C_1k^{-\alpha},\rho'^{(-1)}(c_1)+2C_1k^{-\alpha}] \subset \Gamma_1\,,\\
\xi_2(z,x)\in (\eta_z^\top \nu_k(x)-C_1 k^{-\alpha},\eta_z^\top \nu_k(x))\subset [\rho'^{(-1)}(c_2)-2C_1k^{-\alpha},\rho'^{(-1)}(c_1)+C_1k^{-\alpha}] \subset \Gamma_1\,,
\end{align*}
such that
\begin{align}\label{eqb4}
\begin{aligned}
&\rho'\{\eta_z^\top \nu_k(x)+C_1 k^{-\alpha} \} - \rho'\{ \eta_z^\top \nu_k(x)\}=\rho''\{\xi_1(z,x)\} C_1k^{-\alpha}\geq \inf_{u\in\Gamma_1 } \rho''(u)C_1 k^{-\alpha}\,,\\
&\rho'\{\eta_z^\top \nu_k(x)-C_1 k^{-\alpha} \} - \rho'\{ \eta_z^\top \nu_k(x)\}=-\rho''\{\xi_2(z,x)\} C_1k^{-\alpha}\leq \sup_{u\in\Gamma_1 } \{-\rho''(u)\}C_1 k^{-\alpha}\,,
\end{aligned}
\end{align}
where
\begin{align*}
\Gamma_1=[\rho'^{(-1)}(c_2)-1,\rho'^{(-1)}(c_1)+1].
\end{align*}
Note that $\max\{- \inf_{u\in\Gamma_1 } \rho''(u),\sup_{u\in\Gamma_1 } \{-\rho''(u)\}\}$ is a positive and finite constant, because $\Gamma_1$ is compact and $\rho''$ is continuous and strictly negative. Thus, the claim \eqref{eqb2} is proved using~\eqref{eqb3} and \eqref{eqb4}.

Next, we compute a bound for $\sup_{z\in\mathcal{Z}}\| \nabla H_{hk,z}^*(\eta_z)\|$. We have
\begin{align}\label{eqb5}
&\sup_{z\in\mathcal{Z}} \| \nabla H_{hk,z}^*(\eta_z)\|\notag \\
=\,&\sup_{z\in\mathcal{Z}}\bigg\|\dfrac{E[\rho'\{ \eta_z^\top \nu_{k}(X)\}\nu_k(X) K\{d(Z,z) / h\} ] }{E[ K\{d(Z,z)/ h\}]} -E\{\nu_k(X)\} \bigg\| \notag \\
\leq\,& \sup_{z\in\mathcal{Z}}  \bigg\|\dfrac{E[\rho'\{ \eta_z^\top \nu_{k}(X)\}\nu_k(X) K\{d(Z,z) / h\} ] }{E[ K\{d(Z,z)/ h\}]} - E[ \rho'\{ \eta_z^\top \nu_{k}(X)\}\nu_k(X) |Z=z] \bigg\| \notag\\
&\quad + \sup_{z\in\mathcal{Z}} \big\| E\big( [\rho'\{ \eta_z^\top \nu_{k}(X)\} -\pi_0(z,X)] \nu_k(X)|Z=z \big) \big\| \notag\\
\leq\,&\sqrt{ \sum_{k'=1}^k \sup_{z\in\mathcal{Z}}  \bigg|\dfrac{E[\rho'\{ \eta_z^\top \nu_{k}(X)\}v_{k'}(X) K\{d(Z,z) / h\} ] }{E[ K\{d(Z,z)/ h\}]} - E[ \rho'\{ \eta_z^\top \nu_{k}(X)\}v_{k'}(X) |Z=z] \bigg|^2 }\notag \\
&\quad + \sup_{(z,x)\in \mathcal{Z}\times \mathcal{X}} |\rho'\{\eta_z^\top \nu_k(x)\} - \pi_0(z,x)|  =O(\sqrt{k} h^{\lambda}+k^{-\alpha})\,,
\end{align}
where Lemma~\ref{lm1} and equation \eqref{eqb2} are used for the last inequality.

For all $z\in\mathcal{Z}$ and for some constant $C_2>0$ (to be chosen later), define the set
\begin{align*}
\Lambda_z=\{\eta\in \mathbb{R}^k:\| \eta -\eta_z \|\leq C_2(k^{-\alpha} +h^\lambda \sqrt{k} ) \}\,.
\end{align*}
By Condition~\ref{CB3}, we have $\forall \eta \in \Lambda_z$,
\begin{align*}
\sup_{x\in \mathcal{X}} |\eta^\top \nu_k(x)-\eta_z^\top \nu_k(x)|\leq \|\eta-\eta_z\| \sup_{x\in \mathcal{X}} \|\nu_k(x)\|<C_2(k^{-\alpha}+h^\lambda \sqrt{k})\zeta(k)\,,
\end{align*}
so that $\forall x\in\mathcal{X}$ and for sufficiently large $k$, by~\eqref{eqb2.5},
\begin{align}\label{eqb6}
&\eta^\top \nu_k(x) \notag \\
\in\,& [ \rho'^{(-1)}(c_2)-C_1k^{-\alpha}-C_2(k^{-\alpha}+h^\lambda \sqrt{k})\zeta(k),\rho'^{(-1)}(c_1)+C_1k^{-\alpha}+C_2(k^{-\alpha}+h^\lambda \sqrt{k})\zeta(k)]\subset \Gamma_1\,.
\end{align}

Based on~\eqref{eqb5}, there exists a $C_4>0$ that is independent of $z$ such that $\sup_{z\in \mathcal{Z}}\|\nabla H_{hk,z}^*(\eta_z)\|<C_4(k^{-\alpha}+h^\lambda \sqrt{k})$. For any $\eta \in \partial \Lambda_z$, i.e.~$\| \eta -\eta_z \| = C_2(k^{-\alpha} +h^\lambda \sqrt{k} )$, by the mean value theorem we have 
\begin{align*}
&H_{hk,z}^*(\eta) - H_{hk,z}^* (\eta_z)\\
=\,&(\eta-\eta_z)^\top \nabla H_{hk,z}^*(\eta_z)+\dfrac{1}{2}(\eta-\eta_z)^\top \nabla^2 H_{hk,z}^*(\bar{\eta}_z) (\eta-\eta_z)\\
\leq\,& \|\eta-\eta_z\|\| \nabla H_{hk,z}^*(\eta_z) \|\\
&\quad +\dfrac{1}{2}(\eta-\eta_z)^\top \dfrac{E[ \rho''\{ \bar{\eta}_z^\top \nu_{k}(X)\} K\{d(Z,z) / h\} \nu_k(X)\nu_k(X)^\top ] }{E[ K\{d(Z,z)/ h\}]}  (\eta-\eta_z)\\
\leq\,& \|\eta-\eta_z\|\| \nabla H_{hk,z}^*(\eta_z) \| - \dfrac{a_1}{2}(\eta-\eta_z)^\top { [ E\{\nu_k(X)\nu_k(X)^\top| Z=z\} + O(h^\lambda)J_{k\times k} ] } (\eta-\eta_z)\\
\leq\,& \|\eta-\eta_z\| \Big\{ \| \nabla H_{hk,z}^*(\eta_z) \| - \dfrac{a_1{\{ \lambda_{\rm{min}} +O(h^\lambda k) \} }}{2} \|\eta-\eta_z\| \Big\}\\
\leq\,& \|\eta-\eta_z\| \{C_4(k^{-\alpha}+h^\lambda \sqrt{k}) - a_1{\{\lambda_{\rm{min}}+o(1)\}}C_2(k^{-\alpha}+h^\lambda \sqrt{k})/2 \}\,,
\end{align*}
where $J_{k\times k}$ is the $k\times k$ matrix of ones, $\bar{\eta}_z$ lies between $\eta$ and $\eta_z$ on $\partial \Lambda_z$, $a_1 = \inf_{u\in \Gamma_1}\{-\rho''(u)\}>0$ uniformly in $z\in \mathcal{Z}$, $\lambda_{\min}=\inf_{z\in \mathcal{Z}}\lambda_{z,\min}$ with $\lambda_{z,\min}$ the smallest eigenvalue of $E\{\nu_k(X)\nu_k(X)^\top |Z=z\}$, and we used $O(h^\lambda k)=o(1)$, Lemma~\ref{lm1} and Condition~\ref{CB3}. Therefore, we choose
\begin{align*}
C_2>\dfrac{2 C_4}{a_1 \lambda_{\min}}\,,
\end{align*}
so that $H^*_{hk,z}(\eta)<H^*_{hk,z}(\eta_z)$, for $\eta \in \partial\Lambda_z$. Since $H_{hk,z}^*$ is continuous, there is a local maximum of $H_{hk,z}^*$ in the interior of $\Lambda_z$. On the other hand, $H_{hk,z}^*$ is a strictly concave function with the unique maximum at $\eta_z^*$. Therefore, we conclude that $\eta_z^*\in \Lambda_z^o, \textrm{~i.e.~} \|\eta_z^*-\eta_z\| <C_2(k^{-\alpha}+h^\lambda \sqrt{k})$. Note that $C_2$ is independent of $z$ so that
\begin{align}\label{eqb7}
\sup_{z\in\mathcal{Z}} \|\eta_z^*-\eta_z\| <C_2(k^{-\alpha}+h^\lambda \sqrt{k})\,.
\end{align}

Recalling~\eqref{eqb6} and using~\eqref{eqb7} and Condition~\ref{CB3}, for large enough $k$, we have
\begin{align}\label{eqb8}
&\sup_{(z,x)\in\mathcal{Z}\times\mathcal{X}} |\rho'\{ \eta_z^{*\top}\nu_k(x) \} - \rho'\{\eta_z^{\top}\nu_k(x)\} |\notag \\
=\,&\sup_{(z,x)\in\mathcal{Z}\times\mathcal{X}} | \rho''\{\xi^*(z,x) \} | | \eta_z^{*\top}\nu_k(x) - \eta_z^{\top}\nu_k(x) | \notag\\
\leq\,& a_2 \sup_{z\in\mathcal{Z}} \| \eta_z^*-\eta_z \| \sup_{x\in\mathcal{X}} \|\nu_k(x) \|\leq a_2C_2(k^{-\alpha}+h^\lambda \sqrt{k})\zeta(k)\,,
\end{align}
where $\xi^*(z,x)\in \Gamma_1$ lies between $\eta_z^{*\top}\nu_k(x)$ and $\eta_z^{\top}\nu_k(x)$ and $a_2 = \sup_{u\in\Gamma_1}|\rho''(u)|$. Finally, using~\eqref{eqb2} and \eqref{eqb8}, we conclude that
\begin{align*}
&\sup_{(z,x)\in\mathcal{Z}\times\mathcal{X}}| \pi_{hk}^*(z,x)-\pi_0(z,x) |\\
\leq\,& \sup_{(z,x)\in \mathcal{Z}\times \mathcal{X}}| \pi_{hk}^*(z,x)-\rho'\{\eta_z^\top \nu_k(x)\} |+ \sup_{(z,x)\in\mathcal{Z}\times \mathcal{X}}| \rho'\{\eta_z^\top \nu_k(x)\}-\pi_0(z,x) |\\
=\,& O\{\zeta(k)(k^{-\alpha}+h^\lambda \sqrt{k})\}.
\end{align*}

\end{proof}

\begin{lem}\label{lm3}
Under Conditions~\ref{CB1} to \ref{CB7}, we have
\begin{align*}
\underset{(z,x)\in \mathcal{Z}\times \mathcal{X}}{\sup} |\hat{\pi}_{hk}(z,x)-\pi_{hk}^*(z,x)|=O\bigg\{ \zeta(k)\sqrt{\dfrac{\psi_\mathcal{Z}\{(\log n)/n\}k }{n\mu(h)}} \bigg\}\,.
\end{align*}

\end{lem}

\begin{proof}
The key step is to provide the convergence rate of $\sup_{z\in \mathcal{Z}}\|\hat{\eta}_z-\eta_z^* \|$, for which we need to provide the convergence rate of $\sup_{z\in \mathcal{Z}}\| \nabla H_{hk,z}(\eta_z^*)\|$. For all $z\in\mathcal{Z}$, we write
\begin{align}\label{eqb9}
\hat{S}_n & = \dfrac{\sum_{j\in \mathcal{S}_{-i}} K\{d(Z_j,z)/h\} \nu_k(X_j)\nu_k(X_j)^\top }{\sum_{j\in \mathcal{S}_{-i}} K\{d(Z_j,z)/h\} }  \notag\\
& = \dfrac{E[ K\{d(Z,z)/h\} \nu_k(X)\nu_k(X)^\top] }{E [K\{d(Z,z)/h\}] } + \dfrac{{(n-1)}E[K\{d(Z,z)/h\}]}{ \sum_{j\in \mathcal{S}_{-i}}  K\{d(Z_j,z)/h\}}\notag\\
&\quad \times \bigg[\dfrac{\sum_{j\in \mathcal{S}_{-i}} K\{d(Z_j,z)/h\} \nu_k(X_j)\nu_k(X_j)^\top}{{(n-1)} E[K\{d(Z,z)/h\}]} - \dfrac{E[ K\{d(Z,z)/h\} \nu_k(X)\nu_k(X)^\top ]}{E[K\{d(Z,z)/h\}]} \bigg]\notag\\
&\quad -\bigg(1-\dfrac{{(n-1)}E[K\{d(Z,z)/h\}]}{ \sum_{j\in \mathcal{S}_{-i}} K\{d(Z_j,z)/h\}} \bigg)\dfrac{E[ K\{d(Z,z)/h\} \nu_k(X)\nu_k(X)^\top ]}{E[K\{d(Z,z)/h\}]} \notag\\
&\equiv \dfrac{E[ K\{d(Z,z)/h\} \nu_k(X)\nu_k(X)^\top] }{E [K\{d(Z,z)/h\}] } + \dfrac{{(n-1)} E[K\{d(Z,z)/h\}]}{ \sum_{j\in \mathcal{S}_{-i}}  K\{d(Z_j,z)/h\}} \times \mathcal{A}_{1,z,k}  \notag \\
&\quad-\mathcal{A}_{2,z} \times \dfrac{E[ K\{d(Z,z)/h\} \nu_k(X)\nu_k(X)^\top ]}{E[K\{d(Z,z)/h\}]} \,.
\end{align}
It follows from Lemma~\ref{lm1} that 
\begin{align*}
\sup_{z\in\mathcal{Z}} \mathcal{A}_{1,z,k} =O\bigg\{ \sqrt{\dfrac{ \psi_{\mathcal{Z}}\{(\log n)/n\}}{n\mu(h)}}\bigg\} J_{k\times k} \textrm{ and } \sup_{z\in\mathcal{Z}} \mathcal{A}_{2,z} =O\bigg\{ \sqrt{\dfrac{ \psi_{\mathcal{Z}}\{(\log n)/n\}}{n\mu(h)}}\bigg\} \,.
\end{align*}
Using Lemma~\ref{lm1} again, we have 
\begin{align*}
\sup_{z\in\mathcal{Z}} \dfrac{(n-1)E[K\{d(Z,z)/h\}]}{ \sum_{i\in \mathcal{S}_{-i}}  K\{d(Z_i,z)/h\}}\to 1\,,
\end{align*}
and for any fixed $k$ and all $z\in\mathcal{Z}$
\begin{align*}
\dfrac{E[ K\{d(Z,z)/h\} \nu_k(X)\nu_k(X)^\top ]}{E[K\{d(Z,z)/h\}]}\to E[\nu_k(X)\nu_k(X)^\top|Z=z]\,,
\end{align*}
as $n\to \infty$, whose eigenvalues are bounded away from zero and infinity uniformly in $k$ and $z$ by Condition~\ref{CB3}.

Combing the results above, we deduce that for all $z\in\mathcal{Z}$,
\begin{align}\label{eqb14}
\hat{S}_n=\dfrac{E[ K\{d (Z,z)/h\} \nu_k(X)\nu_k(X)^\top]}{E[K\{d (Z,z)/h\}]}+O\bigg\{ \sqrt{\dfrac{ \psi_{\mathcal{Z}}\{(\log n)/n\}}{n\mu(h)}}\bigg\} J_{k\times k} \,.
\end{align}

By Conditions~\ref{CB3} and \ref{CB5}, for $h$ sufficiently small, there exist two positive constants $s_1$ and $s_2$ that are independent of $z$ such that
\begin{align}\label{eqb15}
0<s_1&\leq \lambda_{\rm{\min}}\bigg(\dfrac{E[K\{d(Z,z)/h\}\nu_k(X)\nu_k(X)^\top]}{E[K\{d(Z,z)/h\}] } \bigg)\notag \\
&\leq \lambda_{\rm{\max}}\bigg(\dfrac{E[K\{d(Z,z)/h\}\nu_k(X)\nu_k(X)^\top]}{E[K\{d(Z,z)/h\}]} \bigg) \leq s_2<\infty\,,
\end{align}
where $\lambda_{\rm{\min}}(A)$ and $\lambda_{\rm{\max}}(A)$ denote the minimum and maximum eigenvalues of $A$, respectively.
Consider the event set for all $z\in\mathcal{Z}$,
\begin{align*}
\mathcal{E}_n(z)=\bigg\{ &(\eta -\eta_z^*)^\top \hat{S}_n (\eta -\eta_z^*)\\
&>(\eta -\eta_z^*)^\top \bigg(\dfrac{E[K\{d(Z,z)/h\}\nu_k(X)\nu_k(X)^\top]}{E[K\{d(Z,z)/h\}] }-\dfrac{s_1}{2}I_{k\times k} \bigg)(\eta -\eta_z^*),~\eta\neq \eta_z^* \bigg\}\,.
\end{align*}
Using~\eqref{eqb14} and \eqref{eqb15}, we deduce that for $h$ sufficiently small,
\begin{align}\label{eqb16}
P\big(\mathcal{E}_n(z)\big) =1 \,.
\end{align}

On the other hand, noting that by defition of $\eta_z^*$,
\begin{align*}
\dfrac{E[ \rho'\{ \eta_z^{*\top} \nu_{k}(X)\} K\{d(Z,z) / h\} \nu_k(X) ]}{E[ K\{d(Z,z)/h\} ] }= E\{\nu_k(X)\}\,,
\end{align*}
we have
\begin{align*}
&\sup_{z\in\mathcal{Z}}\| \nabla H_{hk,z}(\eta_z^*)\|\\
=\,&\sup_{z\in\mathcal{Z}}\bigg\| \dfrac{\sum_{j\in \mathcal{S}_{-i}} \rho'\{ \eta_z^{*\top} \nu_{k}(X_j)\} K\{d(Z_j,z) / h\} \nu_k(X_j)}{\sum_{j\in \mathcal{S}_{-i}}  K\{d(Z_j,z)/h\}} -  \dfrac{1}{n-1} \sum_{j\in \mathcal{S}_{-i}} \nu_k(X_j) \bigg\|\\
\leq\,& \sup_{z\in\mathcal{Z}}\bigg\|\dfrac{1}{n-1} \sum_{j\in \mathcal{S}_{-i}} \dfrac{ \rho'\{ \eta_z^{*\top} \nu_{k}(X_j)\} K\{d(Z_j,z) / h\}\nu_k(X_j) }{E[ K\{d(Z,z)/h\} ] } \\
&\qquad\qquad - \dfrac{E[ \rho'\{ \eta_z^{*\top} \nu_{k}(X)\} K\{d(Z,z) / h\} \nu_k(X) ]}{E[ K\{d(Z,z)/h\} ] } \bigg\| + \bigg\| E\{\nu_k(X)\} -  \dfrac{1}{n-1} \sum_{j\in \mathcal{S}_{-i}} \nu_k(X_j)\bigg\| \\
&\quad +\sup_{z\in\mathcal{Z}}\bigg\| \dfrac{\sum_{j\in \mathcal{S}_{-i}} \rho'\{ \eta_z^{*\top} \nu_{k}(X_j)\} K\{d(Z_j,z ) / h\} \nu_k(X_j)}{\sum_{j\in \mathcal{S}_{-i}} K\{d(Z_j,z)/h\} }\bigg(   \dfrac{\sum_{j\in \mathcal{S}_{-i}} K\{d(Z_j,z)/h\}}{(n-1)E[K\{d(Z,z)/h\}]}-1\bigg)\bigg\| \\
=\,&O\bigg\{\sqrt{\dfrac{\psi_{\mathcal{Z}}\{(\log n)/n\}k}{n\mu(h)}}\bigg\}+O\bigg(\sqrt{\dfrac{k}{n}}\bigg)+O\bigg\{\sqrt{\dfrac{\psi_{\mathcal{Z}}\{(\log n)/n\}k}{n\mu(h)}}\bigg\}=O\bigg\{\sqrt{\dfrac{\psi_{\mathcal{Z}}\{(\log n)/n\}k}{n\mu(h)}}\bigg\}\,,
\end{align*}
where Lemma~\ref{lm1} is used for the second equality. This means that there exists a constant $C_3>0$ that is independent of $z$ such that
\begin{align}\label{eqb17}
 \sup_{z\in \mathcal{Z}}\| \nabla H_{hk,z}(\eta_z^*)\|< C_3 \sqrt{\dfrac{\psi_{\mathcal{Z}}\{(\log n)/n\}k}{n\mu(h)}}\,.
\end{align}

For some $C_4 >0$ (to be chosen later), define the set
\begin{align*}
\hat{\Lambda}_z = \bigg\{\eta\in\mathbb{R}^k:~\|\eta-\eta_z^*\|\leq  C_4 C_3 \sqrt{\dfrac{\psi_{\mathcal{Z}}\{(\log n)/n\}k}{n\mu(h)}} \bigg\}\,.
\end{align*}
For all $\eta \in \hat{\Lambda}_z $, $x\in \mathcal{X}$, and for sufficiently large $n$ and $k$, by Condition~\ref{CB3}, \eqref{eqb2.5} and \eqref{eqb7}, we have
\begin{align}\label{eqb17.3}
&\|\eta^\top \nu_k(x)-\eta_z^{*\top}\nu_k(x)\|\leq \|\eta -\eta_z^{*}\|\|\nu_k(x)\|\leq  C_4 C_3 \sqrt{\dfrac{\psi_{\mathcal{Z}}\{(\log n)/n\}k}{n\mu(h)}} \zeta(k)\notag \\
&\Rightarrow \eta^\top \nu_k(x) \notag \\
&\in \bigg[\eta_z^{*\top}\nu_k(x)- C_4 C_3 \sqrt{\dfrac{\psi_{\mathcal{Z}}\{(\log n)/n\}k}{n\mu(h)}}\zeta(k),\eta_z^{*\top}\nu_k(x)+C_4 C_3 \sqrt{\dfrac{\psi_{\mathcal{Z}}\{(\log n)/n\}k}{n\mu(h)}} \zeta(k) \bigg]\notag \\
&\subset \bigg[\rho'^{(-1)}(c_2)-C_1k^{-\alpha} -C_2(k^{-\alpha}+h^\lambda \sqrt{k}) - C_4 C_3 \sqrt{\dfrac{\psi_{\mathcal{Z}}\{(\log n)/n\} k}{n\mu(h)}} \zeta(k) , \notag\\
&\qquad \rho'^{(-1)}(c_1)+C_1k^{-\alpha} +C_2(k^{-\alpha}+h^\lambda \sqrt{k})+ C_4 C_3 \sqrt{\dfrac{\psi_{\mathcal{Z}}\{(\log n)/n\}k}{n\mu(h)}} \zeta(k)  \bigg] \notag \\
&\subset \Gamma_1 \equiv [\rho'^{(-1)}(c_2)-1,\rho'^{(-1)}(c_1)+1]\,.
\end{align}

By the mean value theorem, for any $\eta \in \partial \hat{\Lambda}_z $, we have, based on the fact that $P(\mathcal{E}_n(z))=1$ from \eqref{eqb16},
\begin{align}\label{eqb17.5}
& H_{hk,z}(\eta) - H_{hk,z}(\eta_z^*)\notag\\
=\,&(\eta-\eta_z^*)^{\top} \nabla H_{hk,z}(\eta_z^*)+\dfrac{1}{2}(\eta-\eta_z^*)^{\top}\nabla^2 H_{hk,z}(\bar{\eta})(\eta-\eta_z^*)\notag\\
\leq\,& \|\eta-\eta_z^* \| \| \nabla H_{hk,z}(\eta_z^*)\|-\dfrac{a}{2}(\eta-\eta_z^*)^{\top} \hat{S}_n (\eta-\eta_z^*)\notag\\
<\,& \|\eta-\eta_z^* \| \|  \nabla H_{hk,z}(\eta_z^*)\|-\dfrac{a}{2}(\eta -\eta_z^*)^\top \bigg[\dfrac{E[K\{d(Z,z)/h\}\nu_k(X)\nu_k(X)^\top\}}{E[K\{d(Z,z)/h\}]}-\dfrac{s_1}{2}I_{k\times k} \bigg](\eta -\eta_z^*)\notag\\
\leq\,& \|\eta-\eta_z^* \| \|   \nabla H_{hk,z}(\eta_z^*)\|-\dfrac{a}{2}(\eta -\eta_z^*)^\top \bigg(s_1 I_{k\times k}-\dfrac{s_1}{2}I_{k\times k} \bigg)(\eta -\eta_z^*)\notag\\
\leq\,& \|\eta-\eta_z^* \|\bigg(\|  \nabla H_{hk,z}(\eta_z^*)\|-\dfrac{a }{4}s_1\|\eta-\eta_z^* \|\bigg)\,,
\end{align}
where $\bar{\eta}\in \partial\hat{\Lambda}_z $ lies between $\eta$ and $\eta_z^*$, $a  = \inf_{u\in \Gamma_1}\{-\rho''(u)\}>0$ and the
third inequality follows from~\eqref{eqb15}. Therefore, we choose
\begin{align*}
C_4>\frac{4}{a s_1}\,,
\end{align*}
so that $H_{hk,z}(\eta)<H_{hk,z}(\eta_z)$, for $\eta \in \partial \hat{\Lambda}_z$. On the other hand, we know that $\hat{\eta}_z$ is the unique maximum point of $H_{hk,z}(\eta)$. It follows that $\hat{\eta}_z \in \hat{\Lambda}_z^o(\epsilon)$, i.e. $\|\hat{\eta}_z-\eta_z^*\|\leq C_4C_3 \sqrt{{\psi_{\mathcal{Z}}\{(\log n)/n\} }k/{n\mu(h)}}$. Noting that $C_4$ and $C_3$ are independent of $z$, we conclude that
\begin{align}\label{eqb20}
\sup_{z\in \mathcal{Z}}\|\hat{\eta}_z-\eta_z^*\|=O\bigg\{ \sqrt{\dfrac{\psi_{\mathcal{Z}}\{(\log n)/n\}k}{n\mu(h)}}\bigg\}\,.
\end{align}

Finally, we are able to prove the convergence rate for $ \sup_{(z,x)\in \mathcal{Z}\times \mathcal{X}} |\hat{\pi}_{hk}(z,x)-\pi_{hk}^*(z,x)|$. By the mean value theorem, we have
\begin{align*}
\hat{\pi}_{hk}(z,x)-\pi_{hk}^*(z,x)=\rho'\{\hat{\eta}_z^\top \nu_k(x)\}-\rho'\{\eta_z^{*\top} \nu_k(x)\}=\rho''\{\bar{\eta}_z^\top \nu_k(x) \}(\hat{\eta}_z -\eta_z^{*})^\top \nu_k(x)\,,
\end{align*}
where $\bar{\eta}_z^\top$ lies between $\hat{\eta}_z$ and $\eta_z^{*}$. From~\eqref{eqb17.3} and~\eqref{eqb20}, we have
\begin{align}\label{eqb21}
\sup_{(z,x)\in \mathcal{Z}\times \mathcal{X}}| \rho''\{\bar{\eta}_z^\top \nu_k(x) \}| =O(1)\,.
\end{align}
It follows that by combining \eqref{eqb20}, \eqref{eqb21} and Condition~\ref{CB3},
\begin{align*}
&\sup_{(z,x)\in \mathcal{Z}\times \mathcal{X}} |\hat{\pi}_{hk}(z,x)-\pi_{hk}^*(z,x)|\\
&\leq \sup_{(z,x) \in \mathcal{Z}\times \mathcal{X}}| \rho''\{\bar{\eta}_z^\top \nu_k(x) \}| \sup_{z\in \mathcal{Z}} \| \hat{\eta}_z -\eta_z^{*}\| \sup_{x\in \mathcal{X}} \|\nu_k(x)\|=O \bigg\{ \zeta(k) \sqrt{\dfrac{\psi_{\mathcal{Z}}\{(\log n)/n\} k}{n\mu(h)}} \bigg\}\,.
\end{align*}

\end{proof}

\section{Proof of Theorem~\ref{theo3}}\label{Appsc4}
To aid the proof of Theorem~\ref{theo3}, we need the following lemma.

\begin{lem}\label{lmd1}
Let $\mathcal{E}_d = \mathcal{E}_d(n)=\{\lambda_d/2\geq \| \hat{G}-G\|_O \}$, where $\|M\|_O=\int \int M^2(s,t)\,ds \,dt$ for a symmetric bivariate function $M$. Under Conditions~\ref{CC1} to \ref{CA2}, we have
\begin{align*}
P(\mathcal{E}_q)\to 1,
\end{align*}
as $n\to \infty$.
\end{lem}
\begin{proof}
This follows from Section 5.1 of~\citet{Hall2007}.
\end{proof}

We give the formal proof of Theorem~\ref{theo3}.

\begin{proof}
From (5.2) of~\citet{Hall2007}, we know that $\sup_{j} |\hat{\lambda}_j-\lambda_j |\leq \|\hat{G}-G\|_O$, where $\|\cdot\|_O$ is defined in Lemma~\ref{lmd1}. Let $\mathcal{E}_q=\{\lambda_q/2\geq \| \hat{G}-G\|_O \}$. We conclude that
\begin{align}\label{eqd1}
\dfrac{1}{2}\lambda_j \leq \hat{\lambda}_j\leq \dfrac{3}{2}\lambda_j\,,
\end{align}
for $j=1,\ldots,q$, provided that the event $\mathcal{E}_q$ holds. We see from Lemma~\ref{lmd1} that $P(\mathcal{E}_q)\to 1$ as $n\to \infty$. Therefore, since our result is probabilistic, we can argue under the assumption that  the event $\mathcal{E}_q$ holds.

Recall that $\hat{b}_\mathrm{FSW}=\sum_{j=1}^q \hat{b}_{\mathrm{FSW},j} \hat{\phi}_j$, where $\hat{b}_{\mathrm{FSW},j}=\hat{\lambda}_j^{-1}\hat{e}_{\mathrm{FSW},j}$. Write $\hat{b}_{\mathrm{FSW},j}=\check{b}_j+\hat{\lambda}_j^{-1}(S_{j1}+S_{j2}+S_{j3})$, where $\check{b}_j=\hat{\lambda}_j^{-1} \int_\mathcal{T}e(t)\phi_j(t)\,dt$, $S_{j1}=\int_\mathcal{T} \{\hat{e}_\mathrm{FSW}(t)-e(t)\}\phi_j(t)\,dt$, $S_{j2}=\int_\mathcal{T} e(t)\{\hat{\phi}_j(t)-\phi_j(t)\}\,dt$ and $S_{j3}=\int_\mathcal{T} \{\hat{e}_\mathrm{FSW}(t)- e(t)\}\{\hat{\phi}_j(t)-\phi_j(t)\}\,dt$. Using~\eqref{eqd1} and the fact that $S_{j3}^2\leq \|\hat{e}_\mathrm{FSW}-e \|^2 \| \hat{\phi}_j-\phi_j\|^2$, we have
\begin{align}\label{eqd2}
\sum_{j=1}^q (\hat{b}_{\mathrm{FSW},j}-\check{b}_j)^2&\leq 3 \sum_{j=1}^q \hat{\lambda}_j^{-2}(S_{j1}^2+S_{j2}^2+S_{j3}^2)\notag \\
&\leq 12 \sum_{j=1}^q  \lambda_j^{-2}(S_{j1}^2+S_{j2}^2+S_{j3}^2)\notag \\
&\leq  12 \sum_{j=1}^q   \lambda_j^{-2}(S_{j1}^2+S_{j2}^2)+12 \sum_{j=1}^q \lambda_j^{-2}\|\hat{e}_\mathrm{FSW}-e \|^2 \| \hat{\phi}_j-\phi_j\|^2\,,
\end{align}

Also, we have
\begin{align}\label{eqd3}
\int_\mathcal{T} \bigg\{\sum_{j=1}^q b_j\hat{\phi}_j(t)-b(t)\bigg\}^2\,dt
&\leq  2\int_\mathcal{T} \bigg[\sum_{j=1}^q b_j\{\hat{\phi}_j(t)-\phi_j(t)\}\bigg]^2\,dt+2\sum_{j=q+1}^\infty b_j^2\notag \\
&\leq 2q\sum_{j=1}^q b_j^2\|\hat{\phi}_j-\phi_j\|^2+2\sum_{j=q+1}^\infty b_j^2 \,.
\end{align}

Using~\eqref{eqd2} and \eqref{eqd3}, we have
\begin{align}\label{eqd4}
&\quad \int_\mathcal{T} \{\hat{b}_\mathrm{FSW}(t)-b(t)\}^2\,dt\notag \\
& =\int_\mathcal{T} \bigg\{\sum_{j=1}^q\hat{b}_{\mathrm{FSW},j}\hat{\phi}_j(t)-\sum_{j=1}^q\check{b}_j\hat{\phi}_j(t)+\sum_{j=1}^q\check{b}_j\hat{\phi}_j(t)-\sum_{j=1}^q b_j\hat{\phi}_j(t)+\sum_{j=1}^q b_j\hat{\phi}_j(t)-b(t) \bigg\}^2\,dt\notag \\
&\leq 3 \sum_{j=1}^q (\hat{b}_{\mathrm{FSW},j}-\check{b}_j)^2+3 \sum_{j=1}^q (\check{b}_j-b_j)^2+3\int_\mathcal{T} \bigg\{\sum_{j=1}^q b_j\hat{\phi}_j(t)-b(t)\bigg\}^2\,dt \notag \\
&\leq 36 \sum_{j=1}^q   \lambda_j^{-2}(S_{j1}^2+S_{j2}^2)+36\sum_{j=1}^q( \lambda_j^{-2}\|\hat{e}_\mathrm{FSW}-e \|^2 +q b_j^2) \| \hat{\phi}_j-\phi_j\|^2+6\sum_{j=q+1}^\infty b_j^2\notag \\
&\qquad+o_p(n^{-(2\beta-1)/(\gamma+2\beta)})\,,
\end{align}
where $\sum_{j=1}^q (\check{b}_j-b_j)^2=o_p(n^{-(2\beta-1)/(\gamma+2\beta)})$ is used for the last inequality. This can be proved following the same argument of proving (5.11) in~\citet{Hall2007} with the response variable $Y$ replaced by $Y \pi_0(Z,X)$. Also, using $q\asymp n^{1/(\gamma+2\beta)}$ and Condition~\ref{CA2}, we have $\sum_{j=q+1}^\infty b_j^2 \asymp \int_{q}^\infty t^{-2\beta}\,dt=O(n^{-(2\beta-1)/(\gamma+2\beta)})$. From Condition~\ref{CA1}, we know that there exists a constant $C_1>1$ such that $\lambda_j\geq C_1j^{-\gamma}$. From (5.21) and (5.22) of~\citet{Hall2007}, we know that $\|\hat{\phi}_j-\phi_j\|^2=O_p(j^2n^{-1})$. Finally, using (5.30) and (5.31) of~\citet{Hall2007}, we have $\sum_{j=1}^q \lambda_j^{-2}S_{j2}^2=O_p(n^{-(2\beta-1)/(\gamma+2\beta)})$. Combining these results with~\eqref{eqd4}, we deduce that there exists a constant $C_2>0$ such that
\begin{align}\label{eqd5}
\int_\mathcal{T} \{\hat{b}_\mathrm{FSW}(t)-b(t)\}^2\,dt \leq C_2 \sum_{j=1}^q \lambda_j^{-2} ( S_{j1}^2 + \|\hat{e}_\mathrm{FSW}-e\|^2 j^2n^{-1})+O_p(n^{-(2\beta-1)/(\gamma+2\beta)})
\end{align}
It remains to provide the convergence rates of $S_{j1}^2$ and $\|\hat{e}_\mathrm{FSW}-e\|^2$.

Recall that
\begin{align*}
\hat{e}_\mathrm{FSW}(t)=\dfrac{1}{n}\sum_{i=1}^n \big\{ \hat{\pi}_{hk}(Z_i,X_i)Y_i-\hat{\mu}_{Y,\pi}\big\} \{Z_i(t)-\hat{\mu}(t)\}\,,
\end{align*}
where $\hat{\mu}_{Y,\pi} = \sum_{i=1}^n \hat{\pi}_{hk}(Z_i,X_i) Y_i/n$. Let
\begin{align*}
\tilde{e}(t)=\dfrac{1}{n}\sum_{i=1}^n \big\{  \pi_{0}(Z_i,X_i)Y_i- \tilde{\mu}_{Y,\pi}\big\} \{Z_i(t)-\hat{\mu}(t)\}\,,
\end{align*}
where $\tilde{\mu}_{Y,\pi} = \sum_{i=1}^n  \pi_{0}(Z_i,X_i) Y_i/n$. Also, recall from the Karhunen-Lo\'eve expansion that $Z_i(t) = \mu(t) + \sum_{j=1}^\infty \xi_{ij} \phi_j(t)$, where the PC scores $\xi_{ij}$ satisfy $E(\xi_{ij})=0$, $E(\xi_{ij}^2)=\lambda_j$ and $E(\xi_{ij}\xi_{ij'})=0$ for $j\neq j'$.

Defining $\bar{\epsilon}=\sum_{i=1}^n\epsilon_i/n$ and $\bar{\xi}_j=\sum_{i=1}^n \xi_{ij}/n$, we can write
\begin{align}\label{eqd50}
S_{j1} &= \int_\mathcal{T} \{\hat{e}_\mathrm{FSW}(t)-\tilde{e}(t)+\tilde{e}(t)-e(t)\}\phi_j(t) \,dt\notag \\
&= \dfrac{1}{n} \sum_{i=1}^n\{Y_i\hat{\pi}_{hk}(Z_i,X_i)-Y_i\pi_0(Z_i,X_i)-\hat{\mu}_{Y,\pi} + \tilde{\mu}_{Y,\pi} \}\int_\mathcal{T} \{Z_i(t) - \hat{\mu}(t)\}\phi_j(t)\,dt \notag \\
&\quad + \dfrac{1}{n} \sum_{i=1}^n\{ Y_i\pi_0(Z_i,X_i)-\tilde{\mu}_{Y,\pi} \} \int_\mathcal{T} \{Z_i(t) - \hat{\mu}(t)\}\phi_j(t)\,dt -E[\{Y\pi_0(Z,X)-\mu_{Y,\pi}\}\xi_{j}] \notag \\
&= \dfrac{1}{n} \sum_{i=1}^n Y_i\{\hat{\pi}_{hk}(Z_i,X_i)-\pi_0(Z_i,X_i)\}(\xi_{ij}-\bar{\xi}_j)\notag \\
&\quad + \dfrac{1}{n} \sum_{i=1}^n \bigg[\int_\mathcal{T}b(t)\{Z_i(t)-\hat{\mu}(t)\}\,dt+\epsilon_i-\bar{\epsilon} \bigg](\xi_{ij}-\bar{\xi}_j)-E[\{Y\pi_0(Z,X)-\mu_{Y,\pi}\} \xi_{j}] \notag \\
&=A_{j1}+A_{j2}+A_{j3}+A_{j4}\,,
\end{align}
where
\begin{align*}
&A_{j1}=\dfrac{1}{n} \sum_{i=1}^n Y_i\{\hat{\pi}_{hk}(Z_i,X_i)-\pi_0(Z_i,X_i)\}(\xi_{ij}-\bar{\xi}_j)\,,\\
&A_{j2}=\dfrac{1}{n} \sum_{i=1}^n \bigg[\xi_{ij}\int_\mathcal{T} b(t)Z_i(t)\,dt-E\bigg\{\xi_{ij}\int_\mathcal{T} b(t)Z_i(t)\,dt\bigg\} \bigg]\,,\\
&A_{j3} = \dfrac{1}{n} \sum_{i=1}^n \{ \xi_{ij}\epsilon_i-E(\xi_{ij}\epsilon_i)\}\,,\\
&A_{j4} = -\bar{\xi}_j\int_\mathcal{T} b(t) \hat{\mu}(t)\,dt-\bar{\xi}_j\bar{\epsilon}\,.
\end{align*}
We next provide the convergence rates for the $A_{jk}$'s sequentially.

First note that $E(\xi_{ij}-\bar{\xi}_j)^2=(n-1) E(\xi_j^2)/n=O(\lambda_j)$ by Condition~\ref{CA1}. It follows that
\begin{align}\label{eqd51}
\dfrac{1}{n}\sum_{i=1}^n (\xi_{ij}-\bar{\xi}_j)^2=O_p(\lambda_j)\,.
\end{align}
Using the Cauchy-Schwarz inequality, we have
\begin{align}
|A_{j1}| &\leq \sup_{(z,x)\in\mathcal{Z}\times \mathcal{X}}| \hat{\pi}_{hk}(z,x)-\pi_0(z,x) | \bigg(\dfrac{1}{n}\sum_{i=1}^nY_i^2\bigg)^{1/2} \bigg(\dfrac{1}{n}\sum_{i=1}^n(\xi_{ij}-\bar{\xi}_j)^2\bigg)^{1/2}  \notag \\
&=O\bigg[\zeta(k)\bigg\{ k^{-\alpha}+h^\lambda \sqrt{k} +  \sqrt{\dfrac{\psi_\mathcal{Z}\{(\log n)/n\} k}{n\mu(h)}} \bigg\}\bigg] \cdot O_p(1)\cdot O_p(\lambda_j^{1/2})\notag \\
&=O_p\bigg[\lambda_j^{1/2}\zeta(k)\bigg\{ k^{-\alpha}+h^\lambda \sqrt{k} +  \sqrt{\dfrac{\psi_\mathcal{Z}\{(\log n)/n\} k}{n\mu(h)}} \bigg\}\bigg] \,,
\end{align}
based on Theorem~\ref{theo2} and \eqref{eqd51}.

For $A_{j2}$, we have
\begin{align*}
\var (A_{j2})\leq \dfrac{1}{n} E\bigg\{\xi_{ij}^2\int_\mathcal{T} b^2(t)Z_i^2(t)\,dt \bigg\}\leq \dfrac{1}{n} \{E(\xi_{j}^4)\}^{1/2}\bigg[E\bigg\{\int_\mathcal{T} b^4(t)Z^4(t)\,dt\bigg\}\bigg]^{1/2}=O\bigg(\dfrac{\lambda_j}{n}\bigg)\,,
\end{align*}
where the last equality follows from Condition~\ref{CA1}. Similarly, we have
\begin{align*}
\var (A_{j3})\leq \dfrac{1}{n}E(\xi_j^2 \epsilon^2)\leq \dfrac{1}{n} \sqrt{E(\xi_j^4)E(\epsilon^4)}=O\bigg(\dfrac{\lambda_j}{n}\bigg)\,,
\end{align*}
where the last equality follows from Conditions~\ref{CA1} and \ref{CA3}. It follows that
\begin{align}\label{eqd6}
A_{j2} = O_p \bigg(\sqrt{\dfrac{\lambda_j}{n}}\bigg)\,,~A_{j3} = O_p \bigg(\sqrt{\dfrac{\lambda_j}{n}}\bigg)\,.
\end{align}

Note that $\hat{\mu}\to \mu$ and $\bar{\epsilon}\to E(\epsilon)$ a.s.~as $n\to \infty$ by the law of large numbers, and $\mu$ and $E(\epsilon)$ are bounded. It follows that $\int_\mathcal{T} b(t)\hat{\mu}(t)\,dt$ and $\bar{\epsilon}$ are bounded for large $n$. Thus, there exists a constant $C_3>0$ such that for $n$ sufficiently large,
\begin{align}\label{eqd7}
|A_{j4}|\leq C_3 \bar{\xi}_j=O_p\Big\{\sqrt{E(\bar{\xi}_j^2)}\Big\}=O_p \bigg(\sqrt{\dfrac{\lambda_j}{n}}\bigg)\,.
\end{align}
Combing~\eqref{eqd50}, \eqref{eqd51}, \eqref{eqd6} and \eqref{eqd7}, we conclude that
\begin{align}\label{eqd8}
S_{j1}^2 = O_p\bigg[\lambda_j\zeta^2(k)\bigg\{ k^{-2\alpha}+h^{2\lambda} k +  \dfrac{\psi_\mathcal{Z}\{(\log n)/n\} k}{n\mu(h)} \bigg\}\bigg]\,.
\end{align}

As for $\|\hat{e}_\mathrm{FSW}-e\|^2$, we have
\begin{align*}
\|\hat{e}_\mathrm{FSW}-e\|^2 \leq 2 \|\hat{e}_\mathrm{FSW}-\tilde{e} \|^2 +2\| \tilde{e}-e\|^2\,,
\end{align*}
where
\begin{align*}
& |\hat{e}_\mathrm{FSW}(t)-\tilde{e}(t)|\\
=\,&\bigg|\dfrac{1}{n} \sum_{i=1}^nY_i\{\hat{\pi}_{hk}(Z_i,X_i)-\pi_0(Z_i,X_i) \} \{Z_i(t) - \hat{\mu}(t)\}\bigg|\\
 \leq\,& \sup_{(z,x)\in\mathcal{Z}\times \mathcal{X}}| \hat{\pi}_{hk}(z,x)-\pi_0(z,x) | \bigg(\dfrac{1}{n} \sum_{i=1}^n Y_i^2 \bigg)^{1/2} \bigg[\dfrac{1}{n} \sum_{i=1}^n \{Z_i(t) - \hat{\mu}(t)\}^2 \bigg]^{1/2}\,.
\end{align*}
It follows by using Theorem~\ref{theo2} and the fact that $\sum_{i=1}^n \int_\mathcal{T}\{Z_i(t) - \hat{\mu}(t)\}^2\,dt /n \to \int_\mathcal{T}\var\{ Z(t)\}\,dt = O(1)$ as $n\to\infty$ that
\begin{align*}
\|\hat{e}_\mathrm{FSW}-\tilde{e} \|^2 =O_p\bigg[\zeta^2(k)\bigg\{ k^{-2\alpha}+h^{2\lambda} k + \dfrac{\psi_\mathcal{Z}\{(\log n)/n\} k}{n\mu(h)} \bigg\} \bigg]\,.
\end{align*}
Also, the standard result on the convergence rate of the empirical mean estimator gives $\|\tilde{e}-e\|^2=O_p(n^{-1})$. Therefore, we conclude that
\begin{align}\label{eqd9}
\|\hat{e}_\mathrm{FSW}-e\|^2=O_p\bigg[ \zeta^2(k)\bigg\{ k^{-2\alpha}+h^{2\lambda} k +  \dfrac{\psi_\mathcal{Z}\{(\log n)/n\}k}{n\mu(h)} \bigg\}\bigg]\,.
\end{align}
The proof is completed by combining~\eqref{eqd5}, \eqref{eqd8} and \eqref{eqd9}.

\end{proof}

\vskip 0.2in
\bibliography{FunTreat_bib}

\end{document}